\documentclass[onecolumn,letterpaper]{aa}
\usepackage{graphicx}
\usepackage{txfonts}
\usepackage{color}

\def\eq{\begin{equation}}         
\def\eeq{\end{equation}}      
\begin{document}
\title{A hot story with many tales: the qWR star \object{HD~45166}
	\thanks{Based on observations made at the 1.5m ESO telescope at La Silla, Chile, and at
	 Laborat\'orio Nacional de Astrof\'isica/CNPq, Brazil}
	}

\subtitle{I. Observations and system parameters}

   \author{J. E. Steiner
         \inst{1}
          \and
          A. S. Oliveira \inst{2}
          }

   \offprints{Alexandre S. Oliveira (\email{aoliveira@ctio.noao.edu})}

   \institute{Instituto de Astronomia, Geof\'{\i}sica e Ci\^encias Atmosf\'ericas (IAG), Universidade de S\~ao Paulo,
             SP, Brasil\\
             \email{steiner@astro.iag.usp.br}
        \and
             SOAR Telescope, Chile\\
            \email{aoliveira@ctio.noao.edu}
             }

   \date{Accepted on 03/07/2005}

   \abstract{The binary star \object{HD~45166} has been observed since 1922 but its orbital period has not
yet been found. It is considered a peculiar Wolf-Rayet star, and its assigned classification varied along the years.
 For this reason we included the object as a
candidate to the class of V~Sge stars and performed spectroscopy in order to search for
its putative orbital period.
      High-resolution spectroscopic observations show that the spectrum, in emission
and in absorption, is quite rich.
The emission
lines have great diversity of widths and profiles. The full widths at half maximum vary
from 70 km s$^{-1}$ for the weakest lines up to 370 km s$^{-1}$ for the most intense ones.
The Hydrogen and Helium lines
are systematically
broader than the CNO lines.
	Assuming that \object{HD~45166} is a double-line spectroscopic binary, it presents an orbital
period of P = 1.596 $\pm$ 0.003 days, with an eccentricity of e = 0.18 $\pm$ 0.08. In
addition, a search for periodicity using standard techniques reveals that the emission
lines present at least two other periods, of 5 hours and of 15 hours. The secondary star
has a spectral type of B7~V and, therefore, should have a mass of about M$_2$ = 4.8 M$_{\sun}$.
Given the radial velocity amplitudes, we determined the mass of the hot (primary) star as being
M$_1$ = 4.2 $\pm$ 0.7 M$_{\sun}$ and the inclination angle of the system, i = 0.77$\degr$ $\pm$ 0.09$\degr$. As the
eccentricity of the orbit is non zero, the Roche lobes increase and decrease as a
function of the orbital phase. At periastron, the secondary star fills its Roche lobe.
	The distance to the star has been re-determined as d = 1.3 $\pm$ 0.2 kpc and a color
excess of $E(B-V)=0.155~\pm~0.007$ has been derived. This implies an absolute $B$ magnitude of $-0.6$ for the primary
 star and $-0.7$ for the B7 star.
	We suggest that the discrete absorption components (DACs) observed in the
ultraviolet with a periodicity similar to the orbital period may be induced by periastron
events.

   }
   \authorrunning{J. E. Steiner \& A. S. Oliveira}
   \titlerunning{\object{HD~45166} -- observations and system parameters}
   \maketitle
%

\section{Introduction}

     	\object{HD~45166} has been observed since 1922, being an enigma that up to now has
not been deciphered. Anger (\cite{ange}) described observations done between 1922 and
1933, showing that the emission spectrum is highly variable. The \ion{He}{ii} 4686 {\AA} line is
always present, with variable intensity, while the lines of \ion{N}{iii} 4640 {\AA} and \ion{N}{iv}
4058 {\AA} appear and disappear completely. The star has been observed ever since, but
no photometric or spectroscopic periodic variation has been discovered, in spite of the
suspicions that it is, in fact, a binary system.

      The spectral classification assigned to this object has been varying along the years. Anger (\cite{ange})
proposed that the object is a Wolf-Rayet of the type WN. Neubauer \& Aller (\cite{neub})
classified it as a W7n. Morgan et al. (\cite{morg}) classified it as Bpe
while Hiltner \& Schild (\cite{hilt2}) recovered the WR classification. Hiltner (\cite{hilt1}) described
the object as having high excitation emission lines superposed upon an approximately O9
spectral type. Heap \& Aller (\cite{heap}), in a frequently cited (see for instance Willis \& Stickland \cite{will1} -- WS83) but
 never published paper,
classified the star as B8~V + qWR, that is, a binary with a B8~V component and a ''quasi''
Wolf-Rayet one. WS83 proposed a classification of B8~V +
SdO -- a main sequence star and a hot sub-dwarf. The Sixth Catalog of Galactic Wolf-
Rayet Stars (van der Hucht et al. \cite{huch1}) recovers for \object{HD~45166} the classification
proposed by Heap \& Aller (\cite{heap}) (B8~V + qWR) and puts it, together with \object{V~Sge}, in the
category of low mass WR, composed by only these two objects. This category does not exist in the Seventh Catalog of
Galactic WR Stars (van der
Hucht \cite{huch2}), since neither \object{HD~45166} nor \object{V~Sge} are included in this catalog.

      The star differs from a classic Population I WR in the sense that its emission lines
are very narrow (typically full width at half maximum (FWHM) = 300 km s$^{-1}$ for the \ion{He}{ii} lines). Besides, the object
presents both characteristics of WN and WC simultaneously and seems to have a
significant abundance of Hydrogen judging from the \ion{He}{ii} Pickering decrement (Heap \& Aller \cite{heap};
WS83). The
object has a much smaller luminosity than a population I WR.

	Van Blerkom (\cite{bler}) made an analysis of the Hydrogen and Helium lines with the
hypothesis that it is a population I WR object. He concluded that the WR component has
a radius of 1 R$_{\sun}$ and has a small size envelope that expands with a velocity of 150 km s$^{-1}$, which results
in number densities of \ion{He}{ii} of about $10^{11}$ cm$^{-3}$ and mimics the environment of a WR envelope. He found that the wind density is $N(H) = 1.6 \times 10^{12}$ cm$^{-3}$ and
$N(He) = 3.0 \times 10^{11}$ cm$^{-3}$, and the mass loss rate, $4.5 \times 10^{-8}$ M$_{\sun}$ year$^{-1}$. On the other hand,
WS83 obtained, on the basis of ultraviolet observations with the IUE satellite, the
following parameters for the WR component: M$_\mathrm{V}=-0.21$; T$_\mathrm{eff}=60\,000$ K; $\log$(L/L$_{\sun}$)=3.84;
R=0.77 R$_{\sun}$ and M$_1$=0.5 M$_{\sun}$. They obtained, also, a distance of 1~208 pc from
spectroscopic parallax of the secondary star and a reddening of $E(B-V)$=0.15 derived
from the 2200 {\AA} interstellar extinction band.

      Willis et al. (\cite{will2}) -- WHSH89 -- based on a sequence of high resolution IUE
spectra, found an extreme constancy of the UV continuum, in contrast to the variability
of the line spectrum. On long time scales they found variability in the intensities of the \ion{C}{iv},
 \ion{N}{iv}, \ion{N}{v} and \ion{He}{ii} emission lines. On short time scales, however, they found
variability exclusively in the \ion{C}{iv} 1559 {\AA} doublet absorption lines. These short time
scale variations ($< 1$ day) are seen as discrete absorption components (DACs)
characterized by two main features, with mean velocities of $-950$ km s$^{-1}$ and $-750$ km s$^{-1}$. DACs
are usually found in stars with P Cygni profiles. They can also be seen in luminous OB
stars as well as in some SdO stars (see Brown et al. \cite{brown} for a recent review on DACs).
In \object{HD~45166}, both components are
observed to migrate in velocity with an acceleration of $\sim 140$ cm s$^{-2}$, a recurrence
time of 1.60 $\pm$ 0.15 days and average lifetime of approximately 3 days. WHSH89
considered that these DACs in \object{HD~45166} could result from structural changes in the
wind, produced by radiative instabilities.

      Until now photometric variability has not been detected in \object{HD~45166}. Ross (\cite{ross})
made $UBV$ observations on 24 nights and did not find any variation. WHSH89
performed 36 hours of UV observations with the IUE satellite and did not find any
variation with amplitude larger than 0.02 magnitudes, that is, within the accuracy limit of
the data. The photometric measurements published in the literature are summarized in
the Table~\ref{photlit}.

There are several properties in common between HD~45166 and the V~Sge stars: Van der Hucht et al. (\cite{huch1})
 already compared the system to \object{V~Sge}. With the parameters proposed by WS83 its mass ratio is inverted 
 as in the V~Sge stars. Moreover, its emission line properties are quite similar. Therefore, we included 
 HD~45166 in our ongoing search program for V~Sge stars.

	The V~Sge stars are a group of 4 stars defined by Steiner \& Diaz (\cite{stei98}): \object{V~Sge}
(Herbig et al. \cite{herb}; Diaz \cite{diaz99}), \object{WX~Cen} (Oliveira \& Steiner \cite{olivwx}; Diaz \& Steiner \cite{diaz95}), \object{V617~Sgr}
(Steiner et al. \cite{stei99}; Cieslinski et al. \cite{cies}) and \object{DI~Cru} (Oliveira et al. \cite{olivdi}; Veen et al. \cite{veen}). They are
characterized by the presence of strong emission lines of \ion{O}{vi} and \ion{N}{v}. Besides, \ion{He}{ii}
4686 {\AA} is at least two times more intense than H$\beta$. The V~Sge stars are very similar
to the Close Binary Supersoft X-rays Sources -- CBSS -- common in Magellanic Clouds, but not that frequent in the Galaxy. The CBSS
are interpreted as binary systems with a white dwarf that presents hydrostatic Hydrogen nuclear burning on its surface.
 This burning is due to the high mass transfer rate which is a consequence of the inverted mass ratio (see Kahabka \& van den Heuvel \cite{kah}, for
a review and references).

      The similarities between the spectra of \object{HD~45166} and of the V~Sge stars are the simultaneous presence
of WN and WC characteristics, strong lines of Hydrogen, as well as the ratios and widths of the emission lines. The main differences
are that \object{HD~45166} presents a spectrum of lower ionization: in spite of having \ion{N}{v}, it doesn't
present \ion{O}{vi}. Besides, the star also presents \ion{He}{i} emission lines, differently
from the V~Sge or CBSS objects. On the other hand, unlike \object{HD~45166}, no V~Sge star or
CBSS has the spectrum of the secondary star published up to now.

      In the next section we will describe our observations. In Sects.~\ref{sectos} and~\ref{b7v} we describe the
optical spectrum of HD~45166 and the spectral classification of the secondary star. In Sect.~\ref{sectop} we derive the orbital period
while in Sect.~\ref{outrper}, two additional periods are proposed. In Sect.~\ref{secmai}, the masses and orbital
inclination are derived. In Sects.~\ref{secd} and~\ref{secc} we present a discussion and conclusions. In
an accompanying paper (Paper II) we study the structure of the wind and discuss the
nature of the system.

\begin{table}[h]
\caption{Photometric magnitudes and colors from the literature.}
\label{photlit}
\begin{flushleft}
\begin{tabular}{lcl}
\hline
\hline\noalign{\smallskip\smallskip}
            \noalign{\smallskip}
Band & Measurement  & Refs. \\  \hline
\noalign{\smallskip}
$V$		&9.983; 9.88	& Menzies et al. (\cite{mara}); Hiltner (\cite{hilt1}) 	\\
$B-V$	&$-0.064;-0.07$		& Menzies et al. (\cite{mara}); Hiltner (\cite{hilt1}) 	\\
$U-B$	&$-0.719;-0.76$		& Menzies et al. (\cite{mara}); Hiltner (\cite{hilt1}) 	\\
$V-R_C$	&0.072			& Menzies et al. (\cite{mara})				\\
$V-I_C$	&0.108			& Menzies et al. (\cite{mara})				\\
$J$		&10.18; 9.81		& Ulla \& Thejll (\cite{ulla}); 2MASS all-sky survey			\\
$H$		&10.12; 9.75	 	& Ulla \& Thejll (\cite{ulla}); 2MASS all-sky survey			\\
$K$		&10.00; 9.57		& Ulla \& Thejll (\cite{ulla}); 2MASS all-sky survey			\\
\noalign{\smallskip}
\hline
\end{tabular}
\end{flushleft}
\smallskip\noindent
\end{table}

\section{Observations}

	 We made spectroscopic observations of \object{HD~45166} from 1998 to 2004 (Table~\ref{jour}), using
the Coud\'e spectrograph at the 1.6 m telescope of Laborat\'orio Nacional de Astrof\'isica (LNA) in Itajub\'a, Brazil, and
the Fiber-fed Extended Range Optical Spectrograph (FEROS) (Kaufer et al. \cite{kauf}) at the 1.52 m
telescope of the European Southern Observatory (ESO) in La Silla, Chile.

At the Coud\'e spectrograph we employed 600 l~mm$^{-1}$ and 1800 l~mm$^{-1}$ gratings,
resulting in spectral resolution of 0.7 and 0.2 {\AA} FWHM and reaching S/N of about 20 to 30 at the continuum.
We used a retro-illuminated Site ($1024 \times 1024$) CCD detector with 24
micrometers resolution elements. Several exposures of bias and flatfield were
obtained to correct for the sensitivity of the CCD. Measurements of dark current were not
necessary. The slit of the spectrograph was adjusted to 250 $\mu$m (about 1.1 arcsec).
Observations of Thorium lamps were made for the purpose of wavelength calibration.
The data reduction was performed with the standard procedures, using IRAF~\footnote
{IRAF is distributed by the National Optical Astronomy Observatories,
which are operated by the Association of Universities for Research in Astronomy, Inc., under cooperative agreement
with the National Science Foundation.}
routines.
Typical calibration $rms$ residuals were of 2 m{\AA} for the 1800 l~mm$^{-1}$ grating observations.

\begin{table*}[h]
\caption{Journal of observations.}
\label{jour}
\begin{flushleft}
\begin{tabular}{lccccc}
\hline
\hline\noalign{\smallskip\smallskip}
            \noalign{\smallskip}
Date (UT)   & Telescope & Spectral                & Exp.          & number   &Spectral\\
                 &                 & resolution ({\AA})   & time (min)   & of exps.  & coverage\\
\hline
\noalign{\smallskip}
1998 Apr 10  & 1.6m LNA    & 0.7    & 2  & 14     &  4520 to 4960 {\AA} \\
1998 Apr 11  & 1.6m LNA    & 0.7    & 2  & 12     &  4520 to 4960 {\AA} \\
1999 Nov 23 & 1.6m LNA    & 0.2    & 10 & 3      &  4590 to 4740 {\AA}  \\
1999 Nov 24 & 1.6m LNA    & 0.2    & 10 & 21    &  4590 to 4740 {\AA}  \\
1999 Nov 26 & 1.6m LNA    & 0.2    & 10 & 18    &  4590 to 4740 {\AA}  \\
1999 Nov 27 & 1.6m LNA    & 0.2    & 10 & 27    &  4590 to 4740 {\AA}  \\
2002 Jan 24 & 1.52m ESO  & 0.1    & 15 & 14    & 3600 to 9200 {\AA}    \\
2002 Jan 25 & 1.52m ESO  & 0.1    & 15 & 10    & 3600 to 9200 {\AA}  \\
2002 Jan 26 & 1.52m ESO  & 0.1    & 15 & 11    & 3600 to 9200 {\AA}   \\
2002 Jan 27 & 1.52m ESO  & 0.1    & 15 & 5      & 3600 to 9200 {\AA} \\
2004 Feb 17 & 1.6m LNA	& 0.7     & 10 & 5      & 4050 to 5180 {\AA} \\
2004 Feb 18 & 1.6m LNA	& 0.7     & 10 & 4      & 4050 to 5180 {\AA}  \\
\noalign{\smallskip}
\hline
\end{tabular}
\end{flushleft}
\smallskip\noindent
\end{table*}

The FEROS spectrograph, on the other hand, uses a bench mounted echelle grating with
reception fibers in the Cassegrain focus. Its measured spectral resolving power is R~=~48~000.
A completely automatic online reduction system was available (Stahl et al. \cite{stahl}) and was adopted
by us. We obtained a total of 40 spectra with integration times of 15 minutes and readout time of about 7 minutes. The signal-to-noise
ratio at the continuum for each individual spectrum was typically S/N=80 at the central wavelength region (4500 through 7000 {\AA}), decreasing to
about S/N=35 at the blue and red edges of the spectrum. The wavelength calibration $rms$ residuals were of 7 m{\AA}. We cut the spectra in
 slices of 250 {\AA} to normalize each slice by interactively fitting a low order Legendre polynomial to the continuum.

\section{The optical spectrum} \label{sectos}

\begin{figure}
   \centering
   \includegraphics[width=10cm]{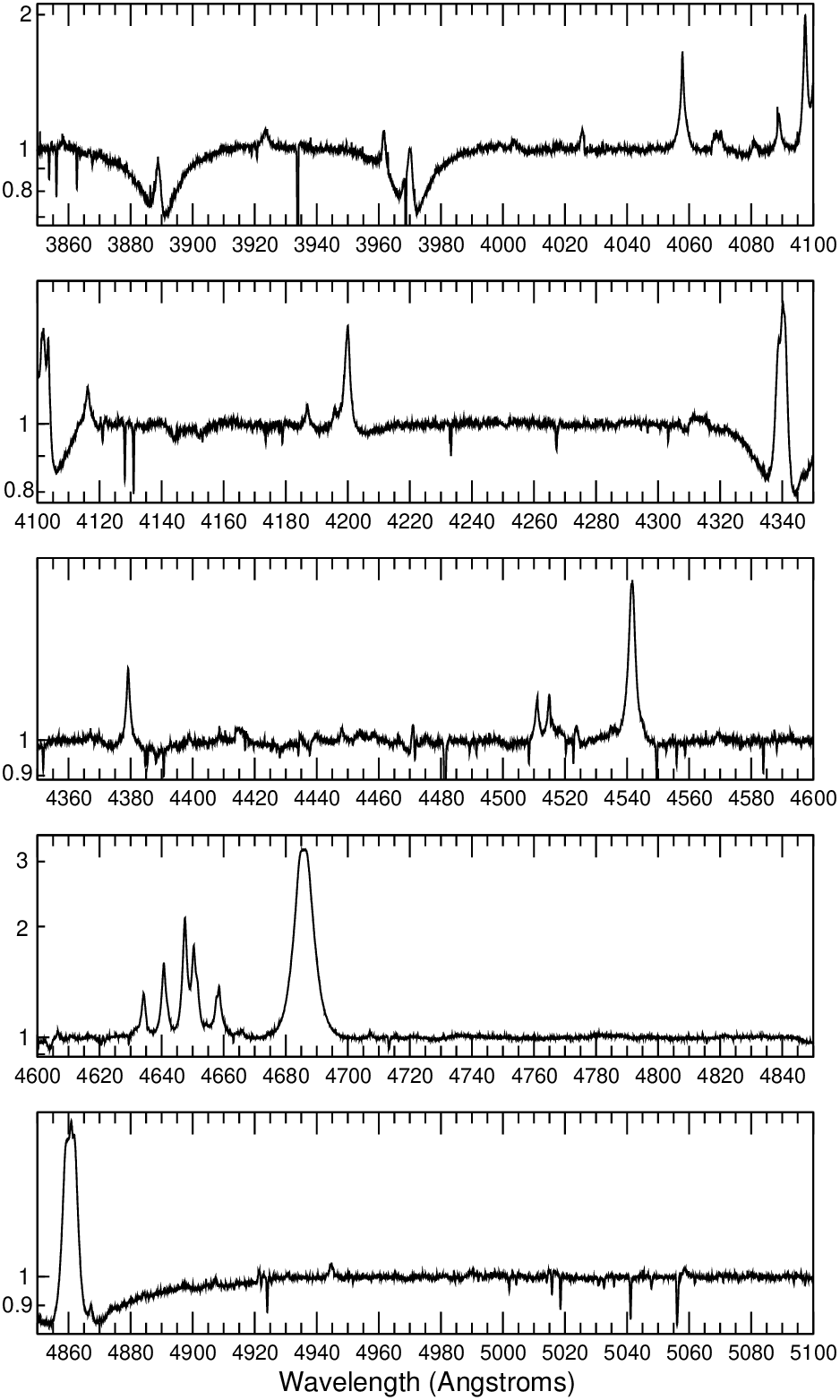}
     \caption{Average normalized spectrum of \object{HD~45166} obtained with
      the FEROS spectrograph. The ordinate is in logarithmic scale.
              }
         \label{sp1}
   \end{figure}

      The average of the 40 spectra obtained with the FEROS spectrograph is presented in Fig.~\ref{sp1} in
logarithmic scale, in order to show a maximum amount of details to the reader.
It presents a large set of lines in absorption as well as in emission (please note the large number of
telluric absorption lines). Quantitative characteristics of the lines are presented
in Tables~\ref{tababs}, \ref{tababs2} and~\ref{tabemis}.
 The spectrum in absorption presents
lines of H, \ion{He}{i}, \ion{C}{ii}, \ion{N}{i}, \ion{O}{i}, \ion{Mg}{ii}, \ion{Si}{ii} and \ion{Fe}{ii}.
Weak lines in absorption of \ion{C}{i}, \ion{Mg}{i}, \ion{Al}{ii}, \ion{Si}{i}, \ion{P}{ii},
\ion{S}{ii}, \ion{Cr}{ii}, \ion{Fe}{i}, and \ion{Ni}{ii} are also present.
The \ion{Fe}{ii} lines present an average heliocentric velocity of $V_r = +6.3 \pm 0.2$ km s$^{-1}$, where the uncertainty
is the standard deviation divided by the square root of the number of measurements.
This $ 0.2$ km s$^{-1}$ uncertainty is 30 times better than the spectral resolution.
The lines of \ion{Si}{ii} have $V_r = +7.3 \pm 0.6$ km s$^{-1}$ and the lines of \ion{O}{i}, $V_r = +5 \pm 1$
km s$^{-1}$. \ion{He}{i} lines present $V_r = +12 \pm 3$ km s$^{-1}$. The
discrepancy between the radial velocities of \ion{He}{i} lines and the other lines can be
explained by the emission that can be seen in the blue wings of some of the \ion{He}{i} lines
as in the case of \ion{He}{i} 4713 {\AA}.


   \begin{figure}
   \centering
   \includegraphics[width=10cm]{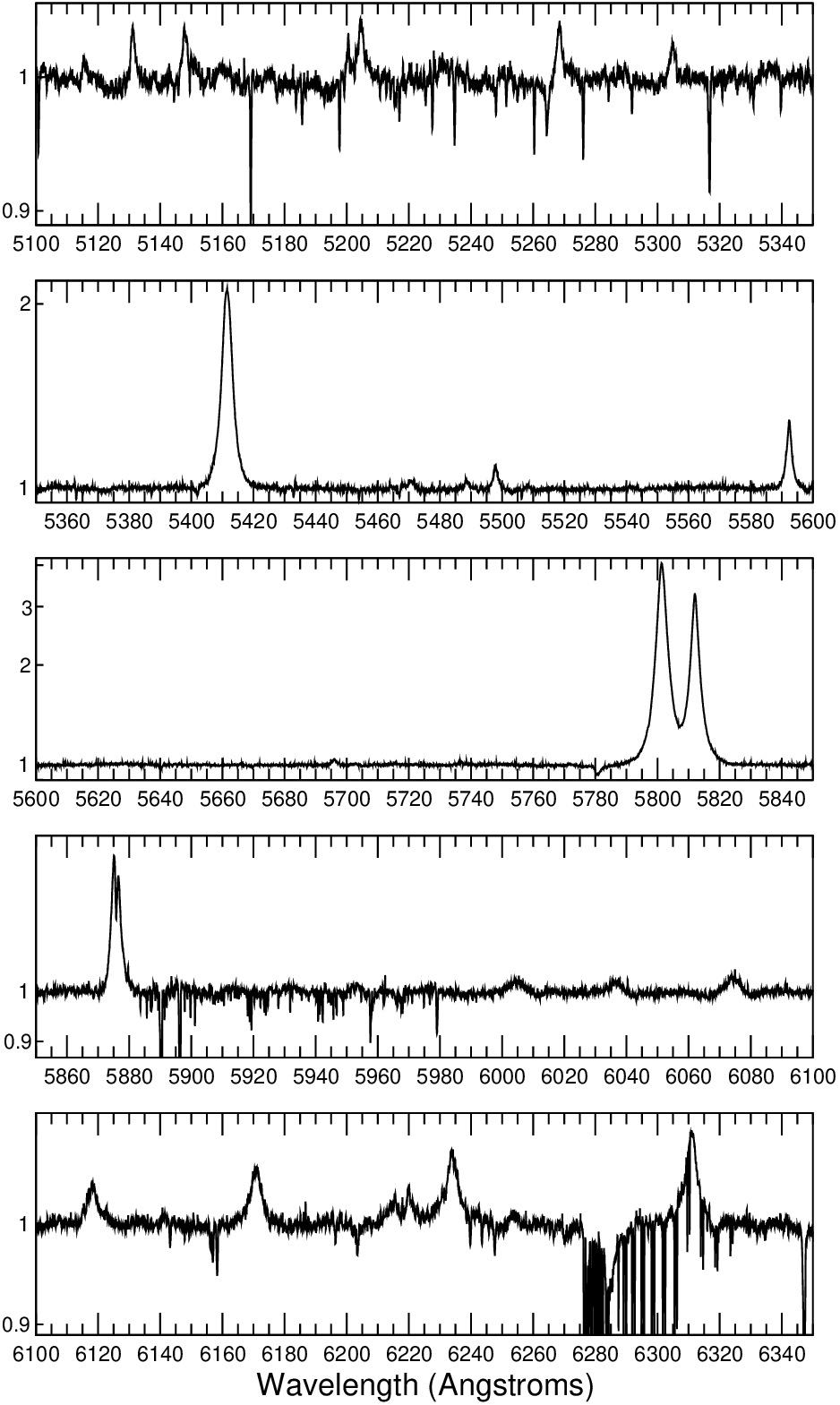}
    \flushleft
    {\bf Fig. 1 (continued).} Average normalized spectrum of \object{HD~45166} obtained with
      the FEROS spectrograph. The ordinate is in logarithmic scale.
         \label{sp2}
   \end{figure}

   \begin{figure}
   \centering
   \includegraphics[width=10cm]{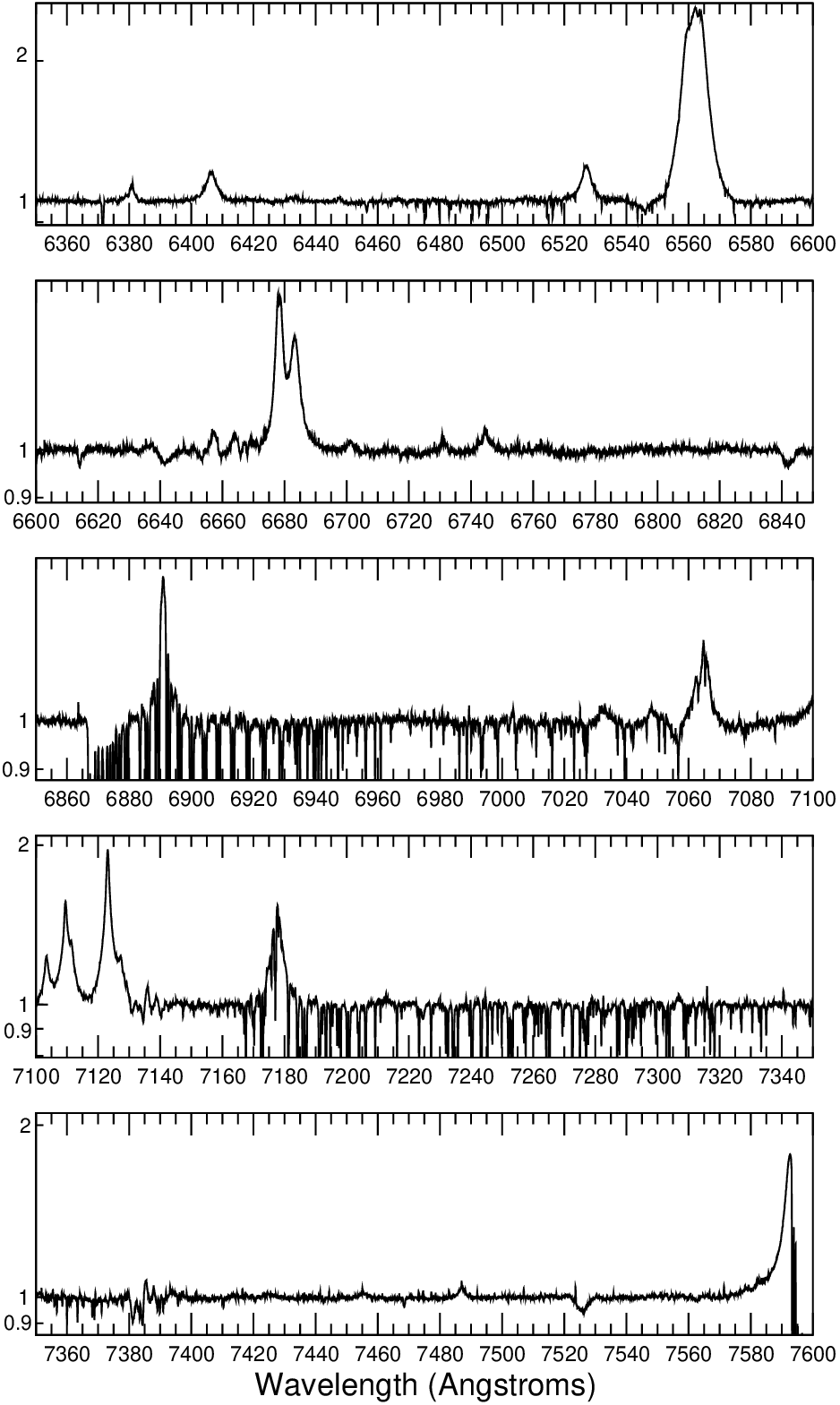}
      \flushleft
    {\bf Fig. 1 (continued).} Average normalized spectrum of \object{HD~45166} obtained with
      the FEROS spectrograph. The ordinate is in logarithmic scale.
         \label{sp3}
   \end{figure}

   \begin{figure}
   \centering
   \includegraphics[width=10cm]{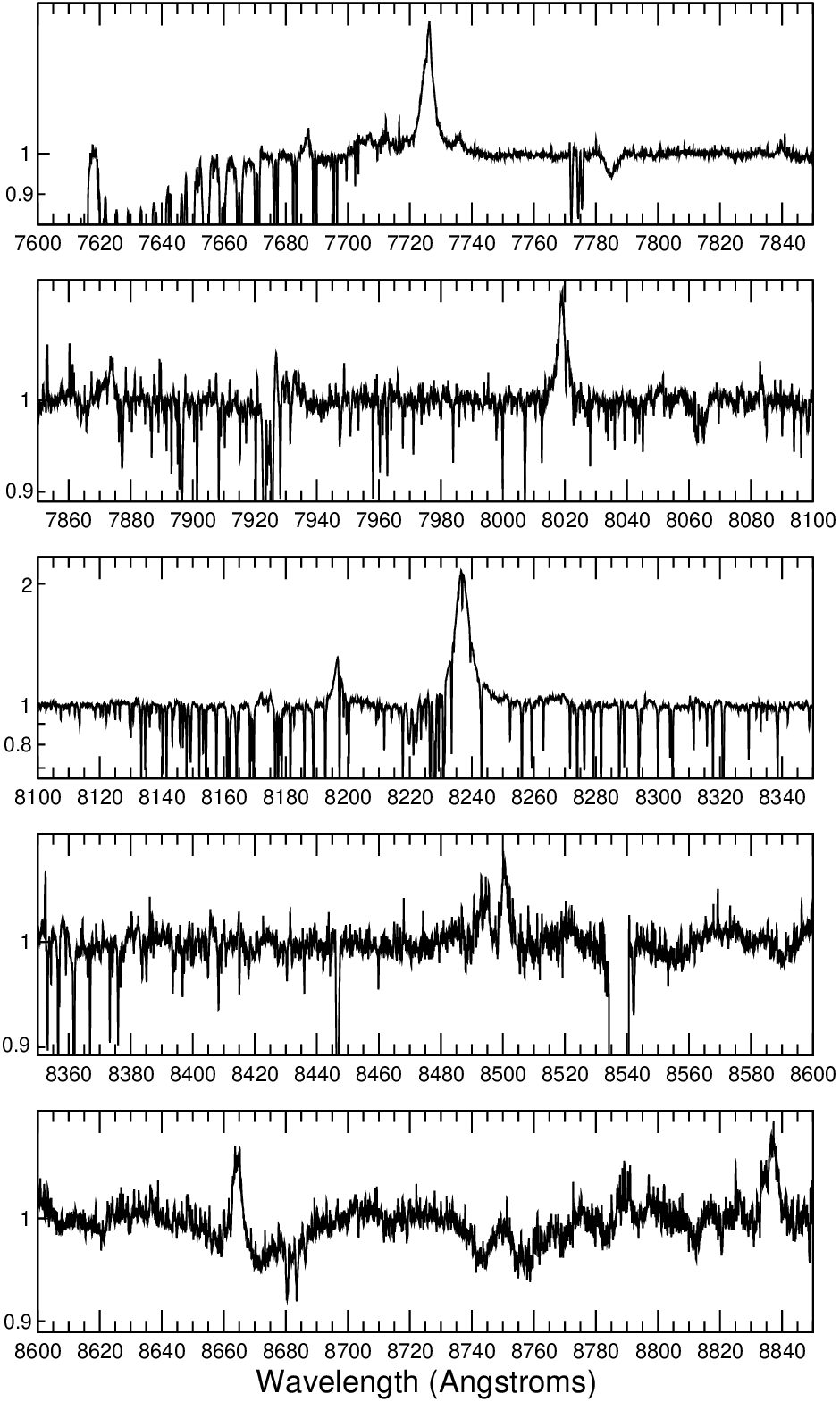}
      \flushleft
    {\bf Fig. 1 (continued).} Average normalized spectrum of \object{HD~45166} obtained with
      the FEROS spectrograph. The ordinate is in logarithmic scale.
         \label{sp4}
   \end{figure}

      A complete list of absorption lines is given in Table~\ref{tababs} for all of the absorption
lines with $W_{\lambda} > 0.02$ {\AA}. Besides the measured wavelength we supply, also, the rest wavelength
(in parenthesis), the radial velocity in the heliocentric system, the equivalent
width, the FWHM and the identification of the species~\footnote{The line identifications were done with the help of the following online
databases: the "NIST Atomic Spectra Database" (http://www.physics.nist.gov/cgi-bin/AtData/main\_asd),
the "Atomic Line List" (http://www.pa.uky.edu/\~{}peter/atomic/) and the "Atomic Molecular and Optical Database Systems"
(http://amods.kaeri.re.kr/spect/SPECT.html).}.
 It is worth to
notice that isolated lines have radial velocities and widths that are comparable amongst
themselves. Doublets, triplets and blended lines are easily identified by the
discrepancies in the values of these measurements. Selected
weak absorption lines are also shown in Table~\ref{tababs2}, where we present only species which have 
not been listed in Table~\ref{tababs}.

\begin{table*}[h]
\caption{Absorption lines with $W_\lambda \geq$~0,02 {\AA}, from the FEROS spectra.}
\label{tababs}
\begin{flushleft}
\begin{tabular}{lllll}
\hline
\hline\noalign{\smallskip}
Wavelength ({\AA})$^{\mathrm{a}}$ & RV (km s$^{-1}$)$^{\mathrm{b}}$ & $W_\lambda$ ({\AA}) & FWHM (km s$^{-1}$) & Identification \\
\hline
\noalign{\smallskip}
3853.76(.67)	&  6.7     &  0.039   &  20.1	 &  \ion{Si}{ii}	\\
3856.10(.02)	&  6.7     &  0.069   &  22.0	 &  \ion{Si}{ii}	\\
3862.69(.59)	&  7.1     &  0.056   &  21.4	 &  \ion{Si}{ii}	\\
4120.99(.81; .82; .99)	&  ...    &  0.028   &  30.9	 &  \ion{He}{i}	\\
4128.15(.05)	&  7.0     &  0.058   &  23.5	 &  \ion{Si}{ii}	\\
4130.98(.87; .89)	&  ...     &  0.072   &  24.8	 &  \ion{Si}{ii}    	\\
4178.96(.86)	&  7.1     &  0.021   &  24.3	 &  \ion{Fe}{ii} 	\\
4233.28(.17)	&  7.8     &  0.033   &  21.7	 &  \ion{Fe}{ii}  	\\
4267.25	&   . . .	   &  0.042   &  33.5	 &  \ion{C}{ii} 	\\
4351.85(.77)	&  5.9     &  0.022   &  21.9	 &  \ion{Fe}{ii}  	\\
4384.73(.64)	&  6.6     &  0.020   &  24.1	 &  \ion{Mg}{ii}+\ion{Fe}{ii}	\\
4481.32(.13; .15; .33)	&  ...     &  0.171   &  34.0	 &  \ion{Mg}{ii}  	\\
4508.37(.29)	&  5.4     &  0.024   &  23.3	 &  \ion{Fe}{ii}  	\\
4522.72(.63)	&  5.6     &  0.026   &  21.8	 &  \ion{Fe}{ii}  	\\
4549.55(.47)	&  4.9     &  0.044   &  27.4	 &  \ion{Fe}{ii}  	\\
4555.99(.89)	&  6.3     &  0.022   &  22.0	 &  \ion{Fe}{ii}  	\\
4583.93(.84)	&  5.9     &  0.032   &  21.6	 &  \ion{Fe}{ii}  	\\
4713.32(.14; .16; .38)	&  . . .	   &  0.037   &  31.8	 &  \ion{He}{i} 	\\
4922.11(1.93)	&  11.2    &  0.023   &  27.4	 &  \ion{He}{i}  	\\
4924.06(3.93)	&  7.7     &  0.047   &  22.9	 &  \ion{Fe}{ii}  	\\
5015.80(.68)	&  19.4    &  0.037   &  25.9	 &  \ion{He}{i}  	\\
5018.55(.44)	&  6.6     &  0.046   &  21.8	 &  \ion{Fe}{ii}  	\\
5041.14(.02)	&  7.0     &  0.058   &  23.5	 &  \ion{Si}{ii}    	\\
5047.86(.74)	&  7.3     &  0.023   &  31.9	 &  \ion{He}{i}  	\\
5056.19(5.98)	&  12.3    &  0.095   &  32.6	 &  \ion{Si}{ii} \\
5100.83(.66; .95)	&  ...	   &  0.025   &  23.5	 &  \ion{Fe}{ii}  	\\
5169.15(.03)	&  6.5     &  0.048   &  21.3	 &  \ion{Fe}{ii}  	\\
5197.68(.58)	&  6.1     &  0.021   &  20.8	 &  \ion{Fe}{ii}  	\\
5227.58(.49)	&  5.3     &  0.025   &  27.8	 &  \ion{Fe}{ii}  	\\
5234.71(.63)	&  4.9     &  0.027   &  26.4	 &  \ion{Fe}{ii}  	\\
5264.45(.22; .36)	&  ...     &  0.029   &  47.3:   &  \ion{Mg}{ii}+\ion{Fe}{ii}  \\
5276.11(.00)	&  6.2     &  0.024   &  21.8	 &  \ion{Fe}{ii}  	\\
5316.73(.62)	&  6.7     &  0.048   &  29.3:   &  \ion{Fe}{ii}  	\\
5640.14(9.98; .24)	&  . . .	   &  0.029   &  54.8	 &  \ion{S}{ii} 	\\
5957.68(.56)	&  6.2     &  0.042   &  23.4	 &  \ion{Si}{ii}    	\\
5979.06(8.93)	&  6.7     &  0.042   &  21.5	 &  \ion{Si}{ii}    	\\
\noalign{\smallskip}
\hline
\end{tabular}
\end{flushleft}
\begin{list}{}{}
\item[$^{\mathrm{a}}$] The values in parenthesis are the rest wavelengths.
\item[$^{\mathrm{b}}$] Velocities in the heliocentric system.
\end{list}
\end{table*}

\begin{table*}[h]
{\bf Table 3.}  Absorption lines with $W_\lambda \geq$~0,02 {\AA}, from the FEROS spectra (continued).
\begin{flushleft}
\begin{tabular}{lllll}
\hline
\hline\noalign{\smallskip}
Wavelength ({\AA})$^{\mathrm{a}}$ & RV (km s$^{-1}$)$^{\mathrm{b}}$ & $W_\lambda$ ({\AA}) & FWHM (km s$^{-1}$) & Identification \\
\hline
\noalign{\smallskip}
6156.93(.74; .76; .78)	&  ...     &  0.027   &  23.1	 &  \ion{O}{i}  	\\
6158.32(.15; .17; .19)	&  ...     &  0.026   &  22.2	 &  \ion{O}{i}  	\\
6347.19(.11; 6.97)	&  ...     &  0.119   &  28.4	 &  \ion{Si}{ii}+ \ion{Mg}{ii}\\
6371.50(.37)	&  6.3     &  0.081   &  23.2	 &  \ion{Si}{ii}    	\\
6402.40		&  . . .	   &  0.023   &   . . . 	 &  \ion{Ne}{i} 	\\
6456.46(.38; 5.84)	&  ...     &  0.031   &  30.1	 &  \ion{Fe}{ii}  	\\
6516.26(.08)	&  8.3     &  0.024   &  11.0:  &  \ion{Fe}{ii}  	\\
7468.47(.31)	&  6.2     &  0.020   &  22.8	 &  \ion{N}{i}  	\\
7772.11(1.99)	&  4.6     &  0.115   &  23.3	 &  \ion{O}{i}  	\\
7774.33(.17)	&  6.4     &  0.133   &  26.2	 &  \ion{O}{i}  	\\
7775.56(.39)	&  6.7     &  0.120   &  27.0	 &  \ion{O}{i}  	\\
7947.51(.17; .55)	&  ...     &  0.034   &  25.1	 &  \ion{O}{i}  	\\
7950.87(.80)	&  2.7    &  0.027   &  23.8	 &  \ion{O}{i}  	\\
8446.70(.25; .36; .76)	&  ...	   &  0.163   &  28.4	 &  \ion{O}{i}  	\\
8680.46(.28)	&  6.2	   &  0.044   &  26.2	 &  \ion{N}{i}  	\\
8683.55(.40)	&  5.1	   &  0.037   &  24.3	 &  \ion{N}{i}  	\\
8686.31(.15)	&  5.6	   &  0.023   &  26.3	 &  \ion{N}{i}  	\\
8719.10(8.84)	&  9.2	   &  0.020   &  22.7	 &  \ion{N}{i}  	\\
 \noalign{\smallskip}
 \multicolumn{5}{c}{Other absorption features} \\
3933.9(.66)	&  18.1    & 0.265   & 28.6    & \ion{Ca}{ii} 		 \\
3968.8(.47)	&  25.1    & 0.128   & 23.5    & \ion{Ca}{ii} 		 \\
4134.5(.11)	&  0.4:	   & 0.032   & 140     & \ion{O}{v}?		\\
4143.96:(.7)	&  20.6:   & 0.113:  & 203:    &  \ion{O}{v}?	\\
4603.91(.74)	&  10.1    & 0.076   & 92      & \ion{N}{v}  	\\
4620.39(9.97)	&  27	   & 0.097   & 200     & \ion{N}{v}  	\\
5890.40(9.95)	&  23.0    & 0.348   & 15.5    & \ion{Na}{i} IS\\
5896.37(5.93)	&  22.5    & 0.282   & 13.3    & \ion{Na}{i} IS\\
5780		&  . . .	   & 0.100	     &  . . .       & DIB\\
\noalign{\smallskip}
\hline
\end{tabular}
\end{flushleft}
\begin{list}{}{}
\item[$^{\mathrm{a}}$] The values in parenthesis are the rest wavelengths.
\item[$^{\mathrm{b}}$] Velocities in the heliocentric system.
\end{list}
\end{table*}

\begin{table*}[h]
\caption{Weak selected absorption lines (0.0035 {\AA} $\leq W_\lambda \leq$ 0,02 {\AA}) from the FEROS spectra.}
\label{tababs2}
\begin{flushleft}
\begin{tabular}{llll}
\hline
\hline\noalign{\smallskip}
Wavelength ({\AA})$^{\mathrm{a}}$ & RV (km s$^{-1}$)$^{\mathrm{b}}$ & $W_\lambda$ ({\AA}) &  Identification \\
\hline
\noalign{\smallskip}
3919.02			&  . . .	 & 0.0129 & 	\ion{C}{ii}+\ion{Fe}{ii} 		\\
4267			&  . . .	 & . . .  & 	\ion{C}{ii} 			\\
4294.50(.40)		&  6.9		 & 0.0066 & 	\ion{S}{ii} 			\\
4588.27(.20)		&  4.5		 & 0.0087 & 	\ion{Cr}{ii} 			\\
4663.15(.05)		&  6.4		 & 0.0157 & 	\ion{Al}{ii}  			\\
4824.21 (.06) 	& 9.5		 & 0.0060 & 	\ion{S}{ii} 			\\
4917.26(.20)		& 3.8		 & 0.0066 & 	\ion{S}{ii} 			\\
4925.46(.34)		& 7.1		 & 0.0066 & 	\ion{S}{ii} 			\\
4992.12(1.97)		& 8.8		 & 0.0089 & 	\ion{S}{ii} 			\\
5032.60(.43)		& 9.7		 & 0.0151 & 	\ion{S}{ii} 			\\
5183.69(.60)		& 5.0		 & 0.0077 & 	\ion{Mg}{i} 			\\
5428.78(.66)		& 7.1		 & 0.0069 & 	\ion{S}{ii} 			\\
5606.27(.15)		& 6.4		 & 0.0061 & 	\ion{S}{ii} 			\\
5852.65(.06)		& 7.4		 & 0.0057 & 	\ion{Si}{i} 			\\
5883.63(.51)		& 6.3		 & 0.0057 & 	\ion{Si}{i} 			\\
5886.96(.83)		& 6.4		 & 0.0057 & 	\ion{Fe}{i}  			\\
5899.73		& ...	 & . . .  & 	\ion{Si}{i}; \ion{Ni}{ii} - (?)	\\
5910.34(.49)		& 7.8		 & 0.0050 & 	\ion{C}{i} 			\\
5912.71(.58)		& 6.3		 & 0.0035 & 	\ion{C}{i} 			\\
5919.36(.20;.23)	& . . . 	 & 0.0110 & 	\ion{Fe}{i}; \ion{S}{ii} 		\\
5940.81(.66)		& 7.5		 & 0.0078 & 	\ion{Cr}{ii} 			\\
5941.35		& 7.8;7.7	 & 0.0076 & 	\ion{Be}{i}; \ion{Cu}{ii} - (?)	\\
5945.72(.51;.54)	& ...	 & 0.0087 & 	\ion{Ni}{ii} 			\\
5948.88		& . . . 	 & 0.0074 & 	\ion{Si}{i}; \ion{Al}{ii} - (?)	\\
5954.67(.47)		& 10.2  	 & 0.0059 & 	\ion{Cr}{ii} \\
5968.01(.91)		&  4.8		 & 0.0050 & 	\ion{P}{ii} \\
5971.07		&  . . .	 & 0.0036 & 	\ion{Cu}{ii}; \ion{C}{i}; \ion{Cr}{i} - (?)	 \\
\noalign{\smallskip}
\hline
\end{tabular}
\end{flushleft}
\begin{list}{}{}
\item[$^{\mathrm{a}}$] The values in parenthesis are the rest wavelengths.
\item[$^{\mathrm{b}}$] Velocities in the heliocentric system.
\end{list}
\end{table*}

	The emission spectrum is also quite rich, presenting lines of H, \ion{He}{i/ii}, \ion{C}{iii/iv}, \ion{N}{iii/iv/v},
 \ion{O}{iii} and \ion{Si}{iv}. In addition, it shows lines that possibly can be identified with \ion{O}{ii}
and \ion{Si}{iii}, as well as some unidentified lines. The H lines are blended with
lines of \ion{He}{ii}, and this affects the measurement of their widths and also their
intensities. In Table~\ref{tabemis} we show a list as complete as possible of all of the emission lines.
Besides the measured and rest wavelengths, we supplied the
radial velocity in the heliocentric system, the equivalent width, FWHM and the identification of the species.

\begin{table*}[!h]
\caption{Emission lines from the FEROS spectra.}
\label{tabemis}
\begin{flushleft}
\begin{tabular}{lllll}
\hline
\hline\noalign{\smallskip\smallskip}
            \noalign{\smallskip}
Wavelength$^{\mathrm{a}}$	& RV	$^{\mathrm{b}}$		& $-W_\lambda$	& FWHM		& Identification	\\
({\AA})		& (km s$^{-1}$)	& ({\AA})			&(km s$^{-1}$)	&			\\
\hline
\noalign{\smallskip}
3702.8:(2.75) 	& 	4.1:      &. . .  &  . . .	&  \ion{O}{iii} \\
3707.33(7.27)	&	4.8   	  & 0.29  & 104     	&  \ion{O}{iii} \\
3715.14(5.09)	&	4.0   	  & 1.0   & 120     	&  \ion{O}{iii} \\
3725.47:(5.94)	&. . . 	          & 0.3   &  . . .	&  \ion{O}{iv}  	\\
3728.85:(9.03)	&. . . 	          & 0.6   &  . . .	&  \ion{O}{iv}  	\\
3754.82(4.70)	&	9.6   	  & 1:    & 128     	&  \ion{O}{iii} \\
3757.3(7.23)	&	6.0   	  & 1:    & 169     	&  \ion{O}{iii} \\
3759.96(9.87)	&	7.2   	  & 1:    & 114     	&  \ion{O}{iii} 	\\
3762.3:		&. . . 	          &. . .  &  . . .	&  \ion{O}{iv}+\ion{Si}{iv}?	\\
3774.11		&. . . 	          &. . .  &  . . .	&  \ion{O}{iv}+\ion{O}{iii}?	\\
3784.3		&. . . 	          &. . .  &   . . .	&  \ion{O}{ii}? \\
3791.2(1.27)	&	-5.5      &. . .  & . . .	&  \ion{O}{iii} \\
3835.1:(5.38)	&. . . 	          &. . .  &  . . .	&  H$\eta$?		\\
3888.80:(8.60)	&. . . 	          & 0.6   &  . . .	&  \ion{He}{i}  	\\
3923.47:(3.48)	&. . . 	      	  & 0.2   & 175     	&  \ion{He}{ii} \\
3961.64(1.57)	&	5.3   	  & 0.3   & 106     	&  \ion{O}{iii} \\
3970.0(9.19)	&. . . 	      	  & 0.8:  & 188     	&  H$\epsilon$+\ion{He}{ii}	\\
4003.6(3.6)	&. . . 	          & 0.06  & . . .	&  \ion{N}{iii} \\
4025.6(5.60)	&. . . 	          & 0.1:  &  . . .	&  \ion{He}{ii} \\
4057.85(7.76)	&	6.6   	  & 1.1   & 103     	&  \ion{N}{iv}  	\\
4069:(8.91)	&. . . 	          &. . .  & . . .	&  \ion{C}{iii} \\
4070.3(0.31)	&. . . 	          &. . .  & . . .	&  \ion{C}{iii} \\
4081.0(1.02)	&	-1.5  	  & 0.15  & 138     	&  \ion{O}{iii} \\
4088.96(8.86)	&	7.3   	  & 0.43  & 81      	&  \ion{Si}{iv} \\
4097.41(7.36)	&	3.7   	  & 0.9   & 88      	&  \ion{N}{iii} \\
4101.8(1.76)	&. . .	      	  & 0.93  & 135     	&  H$\delta$+\ion{He}{ii}		\\
4103.5(3.39)	&	8.0   	  & 0.84  & 82      	&  \ion{N}{iii} \\
4116.2(61)	&	7.1   	  &   0.2 & 102     	 &    \ion{Si}{iv}	\\
4187.0(6.9)	&	7.2	  &   0.1   & 86    	 &    \ion{C}{iii}	\\
4195.82(5.74)	&	5.7	  & . . .   & . . . 	 &    \ion{N}{iii}	\\
4199.97(9.83)	&	10	  &   0.7   & 142   	 &    \ion{He}{ii}	\\
4340.23(0.49)	&. . .		  & . . .   & . . . 	 &    H$\gamma$+\ion{He}{ii}	\\
4379.22(9.11)	&	7.5	  &   0.6   & 97    	 &    \ion{N}{iii}	\\
\noalign{\smallskip}
\hline
\end{tabular}
\end{flushleft}
\begin{list}{}{}
\item[$^{\mathrm{a}}$] The values in parenthesis are the rest wavelengths.
\item[$^{\mathrm{b}}$] Velocities in the heliocentric system.
\end{list}
\end{table*}

\begin{table*}[!h]
{\bf Table 5.} Emission lines from the FEROS spectra (continued).
\begin{flushleft}
\begin{tabular}{lllll}
\hline
\hline\noalign{\smallskip\smallskip}
            \noalign{\smallskip}
Wavelength$^{\mathrm{a}}$	& RV	$^{\mathrm{b}}$		& $-W_\lambda$	& FWHM		& Identification	\\
({\AA})		& (km s$^{-1}$)	& ({\AA})			&(km s$^{-1}$)	&			\\
	    \hline
\noalign{\smallskip}
4448.14:(8.19)	&. . .		&   0.05 	& 88	&    \ion{O}{ii}?	\\
4510.99(0.88)	&	7.3	&   0.3  	& 100	&    \ion{N}{iii}	\\
4514.96(4.85)	&	7.3	&   0.2  	& 73:	&    \ion{N}{iii}	\\
4518.25(8.14)	&	7.3	&   0.1: 	& . . .	&    \ion{N}{iii}	\\
4523.71(3.56)	&	9.9	&   0.1  	& 79	&    \ion{N}{iii}	\\
4534.7(4.58)	&	7.9	& . . .	 	& . . .	&    \ion{N}{iii}	\\
4541.63(1.59)	&	2.6	&   1.65 	& 171	&    \ion{He}{ii}	\\
4569.3 (9.26)	&	2.6	&   0.06 	& 170	&    \ion{O}{iv}	\\
4606.51(6.33)	&	11.7	&   0.07 	& 85	&    \ion{N}{iv}	\\
4610.6(0.55)	&	3.2	& . . .	 	& . . .	&    \ion{N}{iii}	\\
4634.23(4.13)	&	6.5	&   0.75 	& 101	&    \ion{N}{iii}	\\
4640.73(0.64)	&	5.8	&   1.37 	& 100	&    \ion{N}{iii}	\\
4647.51(7.42)	&	5.8	&   2.38 	& 93	&    \ion{C}{iii}	\\
4650.3(0.25)	&	3.2	&   1.46 	& 92	&    \ion{C}{iii}	\\
4651.56(1.47)	&	5.8	& . . .		& . . .	 &    \ion{C}{iii}	\\
4658.36(8.30)	&	3.9	&   0.42:	& 71:	&    \ion{C}{iv}	\\
4665.8(5.86)	&	. . .	&  . . .   	& . . .  &    \ion{C}{iii}	\\
4685.83(5.71)	&	7.7	&     15.6  	&  390  &    \ion{He}{ii}	\\
4707.36(7.31)	&	3.2	&     0.06  	&  108  &    \ion{N}{iv}	\\
4860.59		&	. . .	&     5.3   	&  283  &    H$\beta$+\ion{He}{ii}\\
4867.18(7.17)	&	0.6	&     0.20  	&  90	&    \ion{N}{iii}	\\
4907.2		&	. . .	&     0.04  	&  73	&    \ion{O}{ii}	\\
4944.72(4.56)	&	9.9	&     0.1   	&  109  &    \ion{N}{v}	\\
5058.6(8.73)	&	. . .	&     0.03: 	& . . . &    \ion{N}{iii}	\\
5115.8:(5.41)	&	22.9:	&  . . .    	&  . . .&    \ion{N}{iv}	\\
5131.17		&	. . .	&     0.1   	&  128  &  . . .  	\\
5147.8(7.88)	&	. . .	&     0.08  	&  . . .&    \ion{N}{iii}	\\
5200.5(0.41)	&	5.2	&     0.03: 	& . . . &    \ion{N}{iv}	\\
5204.4(4.28)	&	. . .	&     0.06: 	& . . . &    \ion{N}{iv}	\\
5268.46(8.30)	&	9.1	&     0.1   	&  119  &    \ion{O}{iii}	\\
5305.0(4.97)	&	. . .	&     0.07  	&  . . .&    \ion{Si}{iv}	\\
5411.49(1.52)	&	. . .	&     4.64  	&  216  &    \ion{He}{ii}	\\
5470.3:		&	. . .	&     0.04  	&  . . .&    \ion{C}{iv}	\\
5488.54(8.27)	&	. . .	&     0.04  	&  71	&    \ion{S}{iv}	\\
5497.94(7.78)	&	8.7	&     0.23  	&  98	&    \ion{S}{iv}	\\
\noalign{\smallskip}
\hline
\end{tabular}
\end{flushleft}
\begin{list}{}{}
\item[$^{\mathrm{a}}$] The values in parenthesis are the rest wavelengths.
\item[$^{\mathrm{b}}$] Velocities in the heliocentric system.
\end{list}
\end{table*}

\begin{table*}[!h]
{\bf Table 5.} Emission lines from the FEROS spectra (continued).
\begin{flushleft}
\begin{tabular}{lllll}
\hline
\hline\noalign{\smallskip\smallskip}
            \noalign{\smallskip}
Wavelength$^{\mathrm{a}}$	& RV	$^{\mathrm{b}}$		& $-W_\lambda$	& FWHM		& Identification	\\
({\AA})		& (km s$^{-1}$)	& ({\AA})			&(km s$^{-1}$)	&			\\
	    \hline
\noalign{\smallskip}
5592.40(2.25)	&	8.0	&     0.75  &  96      &    \ion{O}{iii}       \\
5696.0:(5.92)	&	. . .	&     0.09: & . . .    &    \ion{C}{iii}       \\
5737.0:(6.93)	&	. . .	&     0.01: &  . . .   &    \ion{N}{iv}        \\
5801.43(1.31)	&	6.2	&     16.7  &  170     &    \ion{C}{iv}        \\
5812.04(1.97)	&	3.6	&     9.96  &  142     &    \ion{C}{iv}        \\
5875.7(5.6) 	&	. . .	&     1.15  &  189     &    \ion{He}{i}        \\
6004.6:(4.73)	&	. . .	&     0.1:  &  . . .   &       \ion{He}{ii}    \\
6037.0(6.71)	&	. . .	&     0.09: &  . . .   &       \ion{He}{ii}    \\
6074.2 (4.1)    &      	. . .   & 0.3: 	    &	       &	\ion{He}{ii}	       \\
6118.3(8.2)	&	. . .	&     0.2   &  230     &    \ion{He}{ii}       \\
6170.8(0.6)	&	. . .	&     0.3   &  251     &    \ion{He}{ii}       \\
6215.5:(5.45)	&	. . .	&  . . .    &  . . .   &    \ion{N}{iv}        \\
6220.0:(9.89)	&	. . .	&     0.05  &  . . .   &    \ion{N}{iv}        \\
6233.8 (3.8)    &     	. . .	&     0.3   &  . . .   &    \ion{He}{ii}       \\
6310.9(0.8)	&	. . .	&  . . .    &  . . .   &    \ion{He}{ii}       \\
6380.9(0.75)	&	7.0	&  . . .    &  . . .   &    \ion{N}{iv}        \\
6406.4:(6.3)  	&	. . .	&     0.6:  &  . . .   &    \ion{He}{ii}       \\
6527.1(7.11)	&	. . .	&     0.7:  &  179     &    \ion{He}{ii}       \\
6562.0		&	. . .	&     14.4  &  . . .   &    H$\alpha$+\ion{He}{ii}     \\
6678.3(8.15)	&	. . .	&     1.1   &  130     &    \ion{He}{i}        \\
6683.2(3.2)	&	. . .	&     1.0   &  179     &    \ion{He}{ii}       \\
6701.0:		&	. . .	&     0.04: &  . . .   &    \ion{Si}{iv}       \\
6731.1:(0.04)	&	. . .	 & 0.04     &. . .     &  \ion{C}{iii} \\
6744.5(0.39)	&	4.9	 & 0.2	    &182       &  \ion{C}{iii} \\
6890.9:(0.88)	&	. . .	 & . . .    &. . .	       &  \ion{He}{ii} \\
7032.4:(0.34)	&	. . .	 & 0.1	    &. . .     &  \ion{O}{iv}  \\
7064.8:(5.2)	&	. . .	 &. . .	    &. . .     &  \ion{He}{i}  \\
7103.45(0.24)	&	8.9	 & 0.8	    &97        &  \ion{N}{iv}  \\
7109.51(0.35)	&	6.7	 &  1.7	    &97        &  \ion{N}{iv}  \\
7111.6(0.28)	&	13	 & 0.5	    &72        &  \ion{N}{iv}	       \\
7123.1(2.98)	&	5.0	 & 3.1	    &92        &  \ion{N}{iv}	       \\
7127.1(0.25)	&	-6.3	 & 1.1	    &151       &  \ion{N}{iv}	       \\
7177.7(7.50)   	&	. . .	 & . . .    &. . .	       &  \ion{He}{ii} \\
7455.3		&	. . .	 & 0.05     &. . .     &  \ion{O}{iii} \\
7487.0		&	. . .	 & 0.09     &. . .     &  \ion{C}{iii}?        \\
\noalign{\smallskip}
\hline
\end{tabular}
\end{flushleft}
\begin{list}{}{}
\item[$^{\mathrm{a}}$] The values in parenthesis are the rest wavelengths.
\item[$^{\mathrm{b}}$] Velocities in the heliocentric system.
\end{list}
\end{table*}

\begin{table*}[!h]
{\bf Table 5.} Emission lines from the FEROS spectra (continued).
\begin{flushleft}
\begin{tabular}{lllll}
\hline
\hline\noalign{\smallskip\smallskip}
            \noalign{\smallskip}
Wavelength$^{\mathrm{a}}$	& RV	$^{\mathrm{b}}$		& $-W_\lambda$	& FWHM		& Identification	\\
({\AA})		& (km s$^{-1}$)	& ({\AA})			&(km s$^{-1}$)	&			\\
	    \hline
\noalign{\smallskip}
7592(2.74)	&	. . .	&. . . &. . .	&  \ion{He}{ii} \\
7726.2	    	&. . .		&1.3   &112	&  \ion{C}{iv}  	\\
7736.1:		&. . .		&0.1   &. . .	&  \ion{C}{iv}  	\\
7839.8:		&	. . .	&. . . &. . .	&. . .  	\\
7873.7:(3.3)	&	. . .	&. . . &. . .	&  \ion{C}{iii} \\
8019.7 (9.09)   &. . .		&0.5   &. . .	&  \ion{N}{iii} \\
8196.7:(6.48)	&. . .		&. . . &. . .	&  \ion{C}{iii} \\
8237.0:(6.77)	&. . .		&. . . &. . .	&  \ion{He}{ii} \\
8495.0:		&	. . .	&. . . &. . .	&  \ion{N}{iii} \\
8500.5(0.32)	&	. . .	&. . . &. . .	&  \ion{C}{iii} \\
8664.4		&	. . .	&. . . &. . .	&  \ion{C}{iii} \\
\noalign{\smallskip}
\hline
\end{tabular}
\end{flushleft}
\begin{list}{}{}
\item[$^{\mathrm{a}}$] The values in parenthesis are the rest wavelengths.
\item[$^{\mathrm{b}}$] Velocities in the heliocentric system.
\end{list}
\end{table*}

\begin{figure}
   \centering
   \includegraphics[width=10cm]{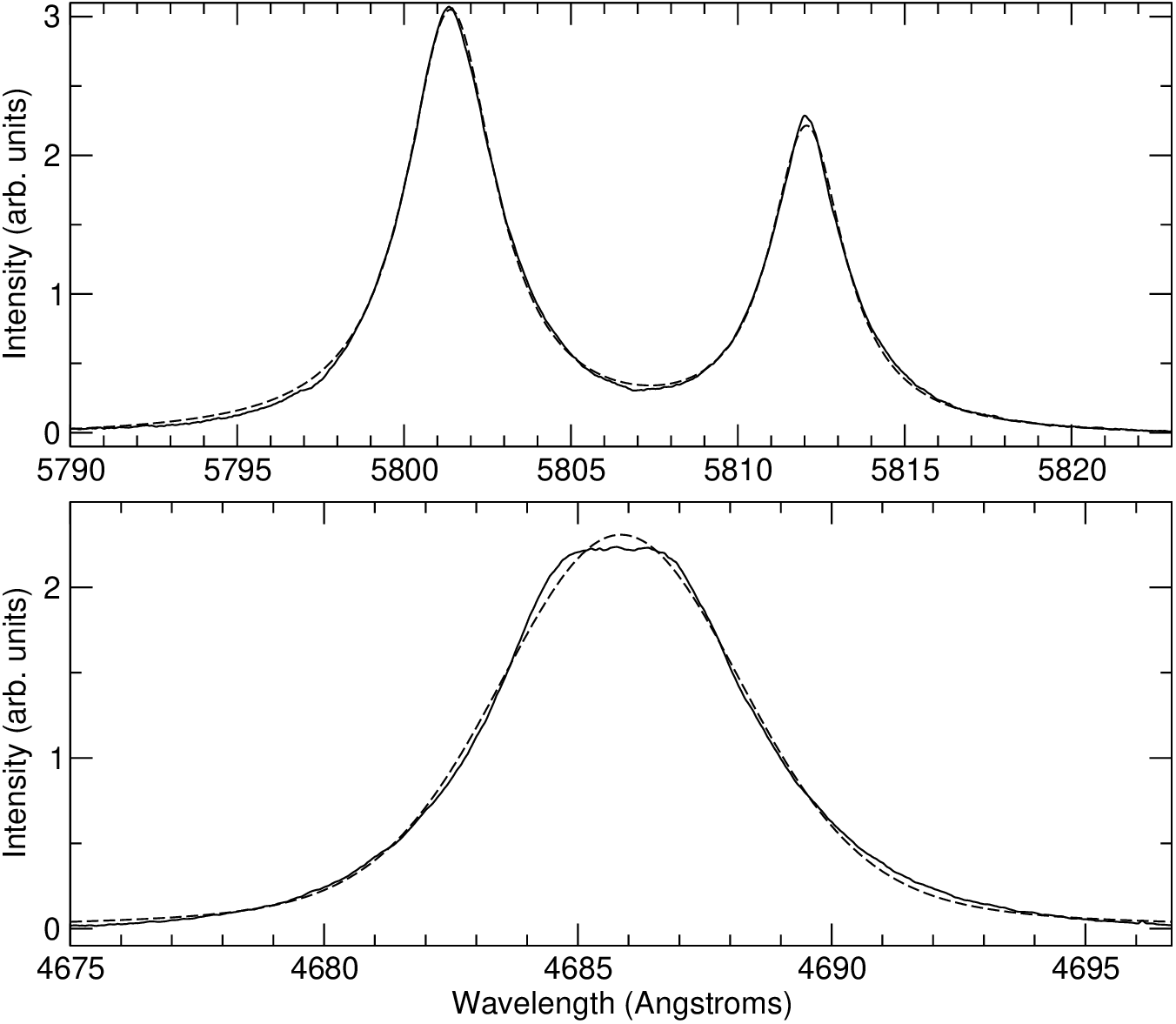}
      \caption{Top: \ion{C}{iv} 5801/11 {\AA} emission lines (continuous line) and Lorentz profiles fitting (dashed line). Bottom:
      \ion{He}{ii} 4686 {\AA} emission line (continuous line) and the Voigt profile fitting (dashed line).
              }
         \label{fit}
   \end{figure}

	An important aspect of the emission line spectrum is the diversity of the widths
and line profiles. The FWHM varies from 70 km s$^{-1}$ for the
weakest lines up to 370 km s$^{-1}$ for the most intense ones. The H and He lines are
systematically broader than the CNO lines and have a profile that can
be described approximately by a profile of Voigt/Gauss, while the CNO lines are quite
well adjusted by a Lorentz profile (see Fig.~\ref{fit}).

      The emission line spectra observed in 1998, 1999, 2002 and 2004 are qualitatively quite similar.
However, the equivalent width of the \ion{He}{ii} 4686 {\AA} line
decreased from $21.0~\pm~0.5$  {\AA} in 1998 and $23.3~\pm~0.4$ {\AA} in 1999 to $15.4~\pm~0.4$ {\AA} in 2002, and
increased again to $22.3~\pm~0.5$ {\AA} in 2004. Still, the FWHM
remained constant along these observations, conserving the value of 380 $\pm$ 10 km~s$^{-1}$.

\section{The spectral classification of the secondary star}\label{b7v}

      The absorption spectrum of the secondary star was detected for the first time by Hiltner (\cite{hilt1})
who described the spectrum of HD~45166 as of ''high excitation superposed on a O9 type star''. Heap \&
Aller (\cite{heap}) classified the secondary as B8~V and this classification has been adopted in
the literature until today. With the high resolution and the high signal-to-noise ratio that
characterize our data, it seems to be adequate to return to the subject and to verify
which, in fact, is the best classification for the star. Many of the conclusions, such as the
values of the masses, depend critically on this parameter.

      In order to re-evaluate the spectral class of the secondary component of the binary system, we compared the
ratio of the observed \ion{Mg}{i} 5183.6 {\AA}/\ion{Mg}{ii} 4481.3 {\AA} equivalent widths
(W$_{\lambda}$(\ion{Mg}{i})/W$_{\lambda}$(\ion{Mg}{ii})~=~$0.0464 \pm 0.0038$)
to
the ones given by the Kurucz (\cite{kurucz})
 models as provided by the VALD (Vienna Atomic Line Database) service
(Kupka et al. \cite{kupka}). We also compared this measured ratio to the ones found at the high resolution
and high S/N ratio spectra of several B type stars from the UVES Paranal Observatory Project
(Bagnulo et al. \cite{bagnu}).
From both comparisons we derived a B7$\pm$1~V spectral type. This classification is
similar, within the errors, to that (B8~V) obtained by Heap \& Aller (\cite{heap}).  According to
Cox (\cite{cox}), the properties of a B7V star are T$_\mathrm{eff}=13~500 $K, M = 4.8 M$_{\sun}$, R = 3.4 R$_{\sun}$ and $M_{B} = -0.7$.

\section{The orbital period} \label{sectop}

      Heap \& Aller \cite{heap} determined that HD~45166 is a binary system composed of a "quasi Wolf-Rayet" star and a
B8 V star.
With the observations described in the previous section, we may now verify whether this
star has or not a detectable orbital period.
If our data present radial velocity variations on both the emission and absorption features than we may be dealing with
a double spectroscopic binary.
However,
previous observations discovered no radial velocity variability in excess of 10 km s$^{-1}$ (WHSH89).
Therefore it is very
important that any measurement shall be made with maximum accuracy.

\begin{figure}
   \centering
   \includegraphics[width=10cm]{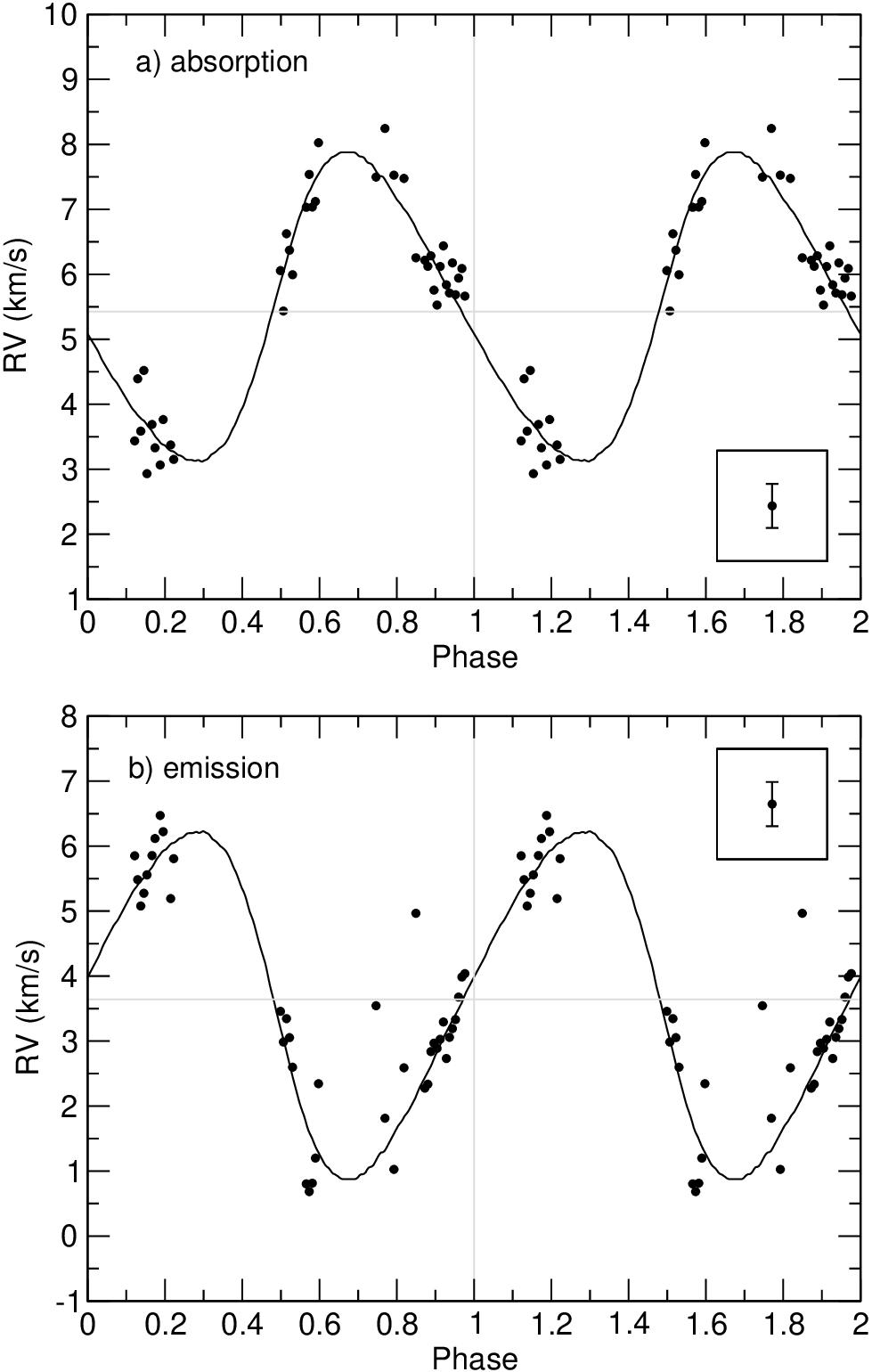}
      \caption{ Radial velocity curves for {\bf a)} \ion{O}{i} absorption lines, as a
      function of the orbital phase, folded to the 1.596 d period.
      {\bf b)} \ion{C}{iv} emission lines, as a
      function of the orbital phase, folded to the same 1.596 d period.
      The 5 and 15 hours modulations (see Sect.~\ref{outrper}) were subtracted from the latter.
      The inset graphics show typical observational error bars.
              }
         \label{perabs}
   \end{figure}

      Photospheric absorption lines of a star are excellent tracers of its movement and potential indicators of an
orbital period. To reach the necessary accuracy, we
measured the radial velocities of the \ion{O}{i} 7772/4/5 and 8446 {\AA} absorption lines from the FEROS spectra and calculated their
weighted average, taking as weight the equivalent widths, respectively W$_\lambda = 0.115$ {\AA} (7772 {\AA}),
W$_\lambda =0.133$ {\AA} (7774 {\AA}), W$_\lambda =0.120$ {\AA} (7775 {\AA}) and W$_\lambda =0.163$ {\AA} (8446 {\AA}).
To proceed, we corrected the \ion{O}{i} average radial velocities for possible uncertainties in the 
wavelength calibration.
 This correction was based on the averaged residuals of the nearby 7670.26 {\AA} and
7671.31 {\AA} telluric lines observed wavelengths, as compared to their rest wavelengths.
The use of telluric absorption lines as stationary comparison source was analyzed by Griffin \& Griffin \cite{griff}.
In our case, to estimate the uncertainties resulting from this correction, one should note that the radial velocities of these 
two telluric lines differ one from the other
by 0.2 km s$^{-1}$ for a same exposure and by 0.5 km s$^{-1}$ for different observations from the same night.
Therefore, the uncertainty in the measurements of the \ion{O}{i} lines position
relative to the nearby telluric features may also be 0.2 km s$^{-1}$, in spite of the observed spectral
 line resolution in this region being 6.2 km s$^{-1}$. 
The relative dispersion of the radial velocities of O I in
Fig.~\ref{perabs} (top diagram) suggests that this is quite reasonable.
One could expect that measurements of absorption lines other than those \ion{O}{i} would have uncertainties of 0.5 km s$^{-1}$,
because of the absence of such correction by close telluric features.

\begin{figure}
   \centering
   \includegraphics[width=10cm]{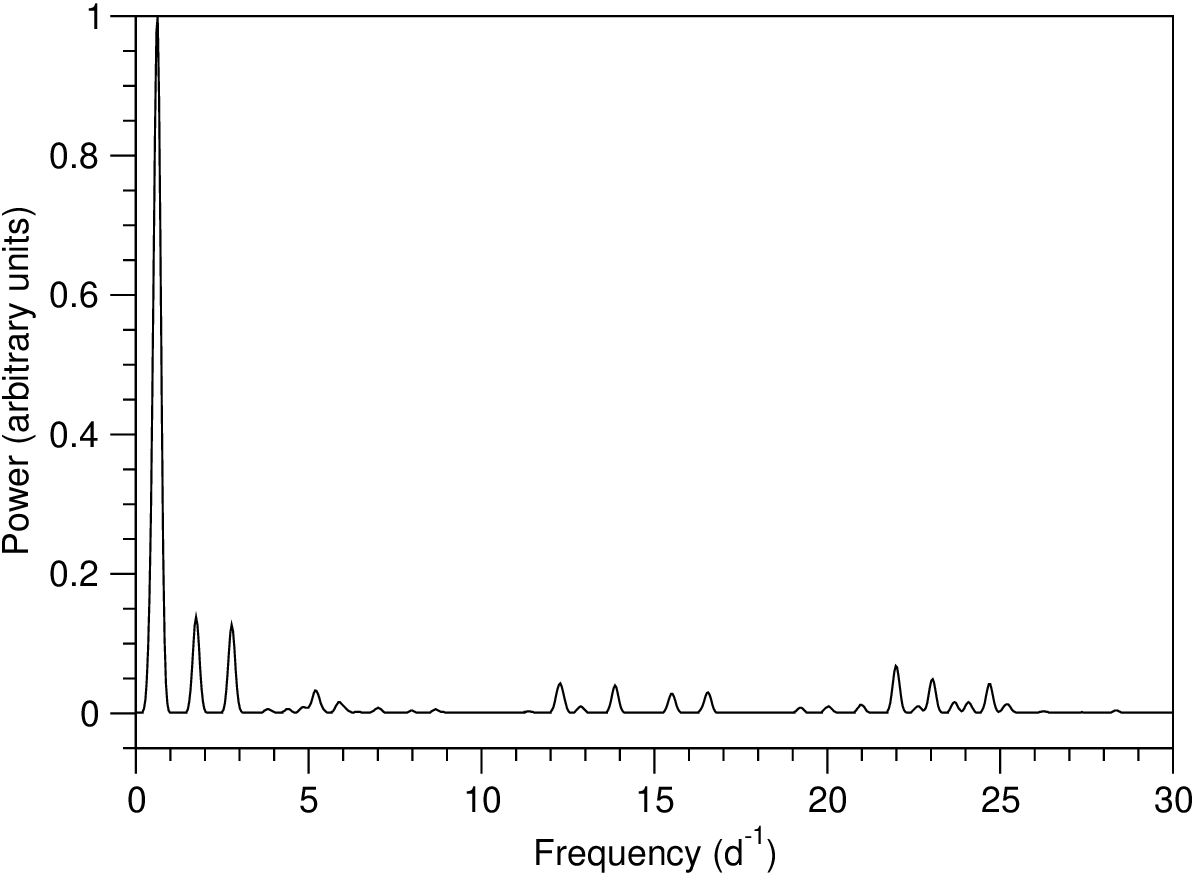}
      \caption{The CLEAN periodogram for the \ion{O}{i} average radial velocities shows a strong peak at frequency f = 0.627 d$^{-1}$, or P = 1.596 d.
              }
         \label{clean}
   \end{figure}

	The CLEAN (Roberts et al. \cite{roberts}) and PDM (Phase Dispersion Minimization, Stellingwerf \cite{stellin})
routines for period search applied to the \ion{O}{i} average radial velocities
indicate a period of P~=~1.596 $\pm$ 0.003 days (Fig.~\ref{clean}). The  associated radial
velocity curve (Fig.~\ref{perabs}, top)
displays an amplitude of K$_2 = 2.4 \pm$ 0.2 km s$^{-1}$ and an eccentricity of e = 0.18 $\pm$ 0.08.
These parameters were obtained fitting the data with the Russell-Wilsing (Binnendijk \cite{binn}) method.

      The mass function is

\eq
f(M_{1})=\frac{M_{1}^{3}{sin}^{3} i}{(M_{1}+M_{2})^{2}} = (1.04\times 10^{-7})(1-e^{2})^{3/2}K_{2}^{3}~P ~~~~~(M_{\sun}) 
\eeq

\noindent where K$_2$ is measured in km s$^{-1}$ and P in days. With the values derived above we obtain a mass 
function of $f(M_1) = 2.2 \times 10^{-6}$~M$_{\sun}$.

\section{The 5 and 15 hours periods} \label{outrper}

	Since the emission spectrum (associated to the primary component of the binary
system) is quite rich, we measured radial displacements of these lines to
verify if the orbital period determined above is also seen in this component. In a first attempt (Oliveira \& Steiner \cite{oliv})
analyzing only the emission lines from the data obtained
at LNA in 1999, we found a candidate period of 0.357 days.

\begin{figure}
   \centering
   \includegraphics[width=10cm]{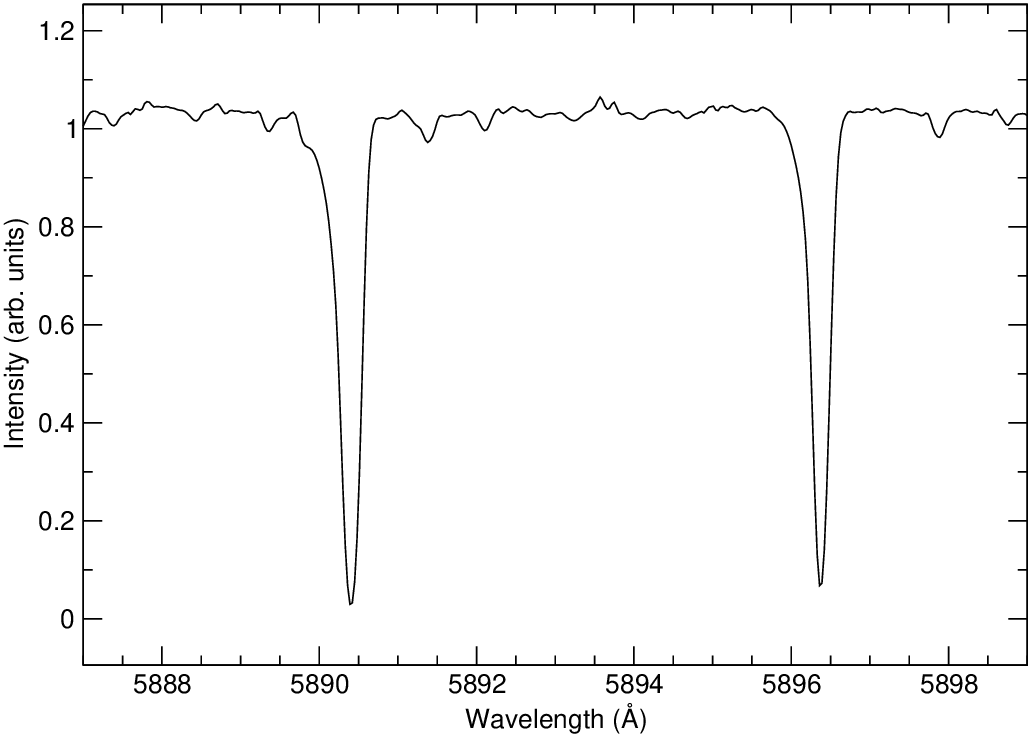}
      \caption{The \ion{Na}{i} interstellar absorption profiles, obtained with the FEROS spectrograph.
              }
         \label{nafig}
   \end{figure}

To measure the radial velocities of the emission lines from the FEROS spectra we used the \ion{C}{iv} 5801/11 {\AA}
lines, because these features are narrower than the H and He lines
and they are very well fitted by Lorentzian profiles (Fig.~\ref{fit}, top diagram). Besides, they are located only 100 {\AA} away from
the \ion{Na}{i} 5890/96 {\AA} interstellar absorption lines (Fig.~\ref{nafig}), which can be used as
fiducial marks to correct for radial velocity inaccuracies. This is the same procedure we applied to the \ion{O}{i} absorption lines above,
when the fiducial references were telluric features.
Therefore, the \ion{C}{iv} radial velocity measurements were performed by simultaneously fitting Lorentz profiles
and obtaining the simple
average of both radial velocities. Then we subtracted, from this average, the residuals of the radial
velocities of the \ion{Na}{i} interstellar lines relative to their
median value.

\begin{figure}
   \centering
   \includegraphics[width=10cm]{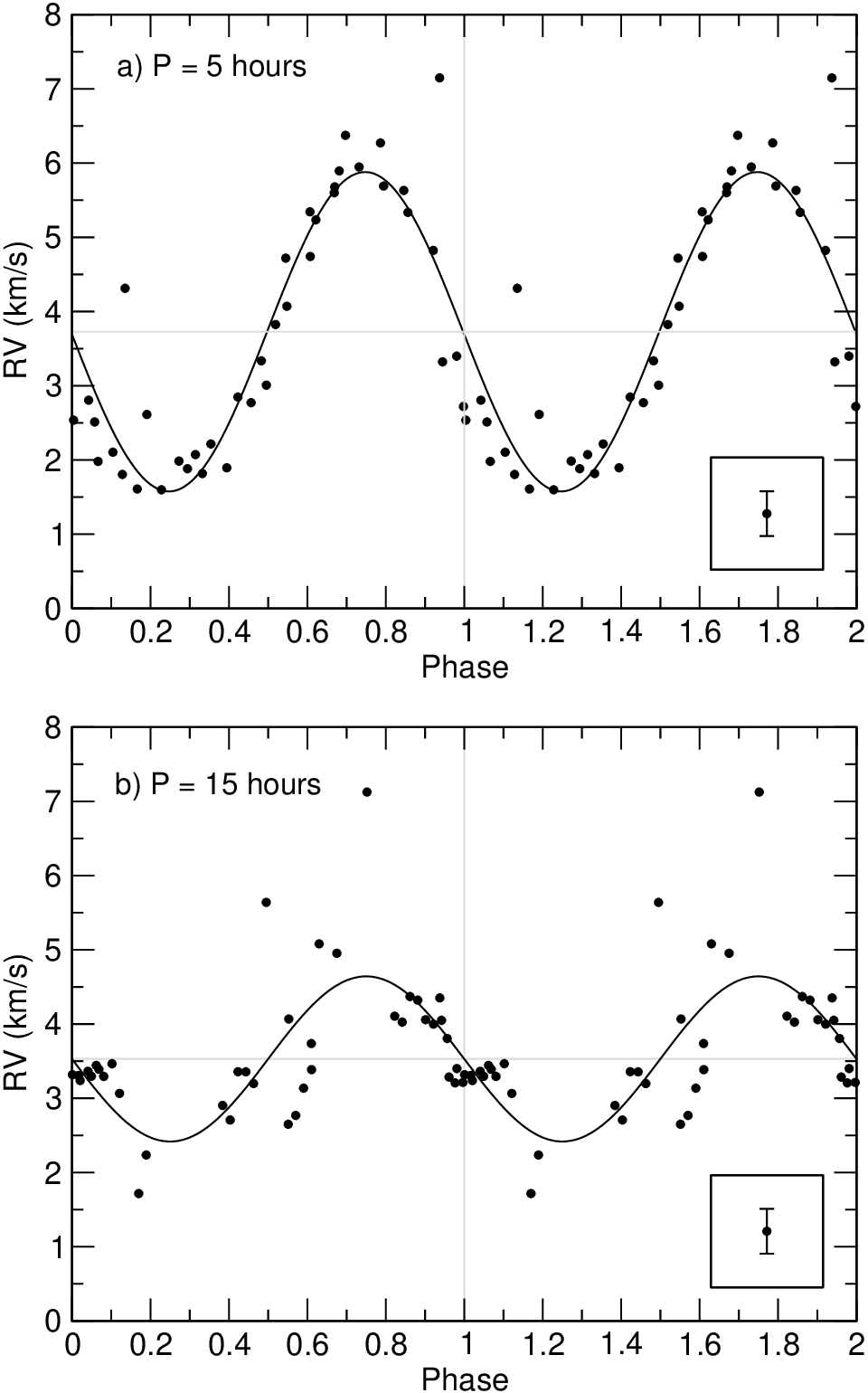}
      \caption{Radial velocity curves for the \ion{C}{iv} emission lines,
      as a function of {\bf a)} the 5 hours period (the 1.596 days and 15 hours modulations were previously subtracted) and
      {\bf b)} the 15 hours period (the 1.596 days and 5 hours modulations were previously subtracted). The inset graphics show
      typical observational error bars.
              }
         \label{peremis}
   \end{figure}

      The resultant radial velocity curve shows a period of 0.205 days
(5 hours) (Fig.~\ref{peremis}, top). The fact that two spectroscopic
periods exist is a complicating factor because, with the small amplitudes involved, it is
very difficult to determine them.
Another factor that complicates the determination of the periods is that their
values coincide with unit fractions of a day. We have confused (Oliveira \& Steiner \cite{oliv})
the period of 0.205 days ($\sim 1/5$ of a day) with 0.357 days ($\sim 1/3$ of a day). In the
periodogram obtained from the LNA data, the most visible period is, in fact, $\sim 1/3$ of a
day. One should mention that the orbital period is $\sim 3/2$ of a day.

	At this time we should wonder which, after all, is the orbital period as we have now identified two periods.
	 We believe that the 1.596 days period is orbital for two
strong reasons. First, it was derived using the photospheric absorption lines from the
secondary star.  Second, the 5 hours period is so short that, if it were
orbital, the secondary star would not fit in its Roche lobe.

      If the 1.596 $\pm$ 0.003 days period is the orbital one, we should find this period
also in the primary star, that is, we should search for this period in the emission lines. To do
this, we subtracted the average radial velocity curve, folded with the
5 hours period, from the raw data of \ion{C}{iv} radial velocity. The result is plotted as a function
of the 1.596 days period (Fig.~\ref{perabs}, bottom) and shows a curve symmetric to the absorption radial velocity curve
with the same period, but with an amplitude slightly larger, that is,
K$_1 = 2.7 \pm$ 0.2 km s$^{-1}$. 
The radial velocity curve derived from the \ion{C}{iv} emission lines is in anti-phase with the \ion{O}{i} absorption lines curve. This should be expected if the emission and absorption lines are produced on distinct stellar components of the binary system and if the sample period is the orbital one. To investigate whether the \ion{C}{iv} radial velocity variations could be attributed to superposed photospheric absorption features from the secondary component, we searched for \ion{C}{iv} 5801/11 {\AA} absorption lines in high resolution spectra of B stars. Our FEROS spectra of HD~103401 (B5 V), 
HD~104432 and HD~102465 (both B9 V), taken with the same instrumental configuration as the spectra of HD~45166, do not show any features above the noise that could be associated to \ion{C}{iv} 5801/11 {\AA} lines. This fact, together with the anti-phase behavior of the radial velocity curves, lead to the interpretation of the \ion{C}{iv} radial velocity variations as produced by the orbital motion of the primary star.

The mass function of the secondary component is

\eq
f(M_{2}) = 3.1 \times 10^{-6}M_{\sun}
\eeq

	After we subtract the 1.596 days and the 5 hours modulations, we applied the PDM
routine on the residuals and we identified an additional
period of 0.64 days (15 hours, Fig.~\ref{peremis}, bottom).
Using these techniques, no other candidate period
appeared with amplitude greater than 0.5 km s$^{-1}$. In Paper II we will return to the non-orbital
 periods and their interpretation.

 \section{Masses and inclination} \label{secmai}

      The orbital period was detected in both stars (associated to the absorption and
also emission lines), so this system is a double spectroscopic binary.
With the radial velocity amplitudes from both stars, we obtain

\eq
q = \frac{M_{2}}{M_{1}} = \frac{K_{1}}{K_{2}} = 1.13 ~{\pm}~ 0.11
\eeq

As a main sequence B7~V star, the mass of the secondary is M$_2 = 4.8~{\pm}~0.5$ M$_{\sun}$;
therefore M$_1$ = 4.2 $\pm$ 0.7 M$_{\sun}$. The mass of the primary star was estimated by
WS83. They assume that the primary star obeys the wind relations of typical OB stars.
From an approximate wind model and observed P-Cygni profile parameters from IUE, they estimate a mass 
of 0.5 M$_{\sun}$. In that case the mass was not inferred from dynamical parameters such as radial velocity curves.

The inclination angle of the system, obtained from
the mass function, is i = 0.77$\degr$ $\pm$ 0.09$\degr$. This is probably one of the smallest orbital
inclination angles known for a binary system. 

	The major semi-axes of the orbits, (a$_1$,a$_2$), are defined as

\eq
a_{1,2}~ sin~ i = (1.98 \times 10^{-2})(1-e^{2})^{1/2}K_{1,2}~P~~~~~R_{\sun} 
\eeq

\noindent and, in the case of \object{HD~45166}, are a$_1 = 6.3 \pm$ 0.6 R$_{\sun}$ and a$_2 = 5.5 \pm$ 0.5 R$_{\sun}$.

	Would it be the case to ask whether the secondary star fills its Roche lobe? As
the eccentricity of the orbit is non zero, the Roche lobes increase and decrease as a
function of the orbital phase. At periastron, the Roche lobes have minimum dimensions
and, at that phase, the effective radius of the secondary's lobe is given by (Hilditch \cite{hildi})

\eq
R_{r2}(min)= \frac{4.21~P^{2/3}(1-e)(K_{1}+K_{2})(0.38+0.20~ \log (M_{2}/M_{1}))M_{1}}
{K_{2}(M_{1}+M_{2})^{2/3}}~~~~~R_{\sun}
\eeq

For the values determined for \object{HD~45166}, we have $R_{r2}(min)$ = 3.7 ~${\pm}$~0.8 $R_{\sun}$.
Considering the uncertainties, this value is very close to the secondary's radius (see Sect.~\ref{b7v}) and, therefore,
 at periastron the secondary star may fill its Roche lobe.

       The maximum radius for the secondary's Roche lobe is (Hilditch \cite{hildi})

\eq
R_{r2}(max)= \frac{4.21~P^{2/3}(1+e)(K_{1}+K_{2})(0.38+0.20~ \log (M_{2}/M_{1}))M_{1}} 
{K_{2}(M_{1}+M_{2})^{2/3}}~~~~~R_{\sun}
\eeq

\noindent that is, 5.4 $\pm$ 1.1 R$_{\sun}$. In Table~\ref{parsis} we list the basic parameters for the system.

\section{Discussion} \label{secd}

\subsection{Gravitational redshift}

	The General Theory of Relativity predicts the existence of gravitational redshift
given by (for $z = v/c \ll 1$, where $z$ is the redshift, $v$ is velocity and $c$ is the velocity of light)

\eq
z= \frac{V_\mathrm{gr}}{c} = \frac{GM}{R~c^{2}}=\frac{2.11\times 10^{-6} M/M_{\sun}}{R/R_{\sun}}
\eeq

      At the surface of the secondary star this corresponds to $V_\mathrm{gr}$  = +0.9 km s$^{-1}$ and for
the primary star, $V_\mathrm{gr}$ = +2.0 km s$^{-1}$ (where we considered $R_{1}=1.3~R_{\sun}$, see Paper II).
 As these values are larger than the precision of our
measurements, they need to be taken into account in the data analysis. If the average
velocity of the secondary absorption spectrum is +6.2 $\pm$ 0.3 km s$^{-1}$, then the expected
average photospheric velocity in our spectra of the primary star should be of +7.3 km s$^{-1}$.
The average heliocentric velocity of the secondary star, corrected for the gravitational
redshift, is $<v(HC)>$ = +5.3 ~${\pm}$~ 0.3 km s$^{-1}$. This velocity is not the radial velocity of the
center of mass, because this depends on the sampling of the observation phases.
 The value of $\gamma$, that is, of the radial velocity of the mass center, is
determined from the radial velocity curve and corresponds to the value that divides the
curve in two segments of same underlying areas. Its value, obtained from the secondary star's absorption lines radial
velocities and corrected for the gravitational redshift, is $\gamma$= 4.5 ~${\pm}$~0.2 km s$^{-1}$ (Table~\ref{parsis}).

\begin{table*}[!h]
\caption{Basic parameters of the binary system.}
\label{parsis}
\begin{flushleft}
\begin{tabular}{lll}
\hline
\hline\noalign{\smallskip}
\multicolumn{3}{c}{Parameters of the system} \\
\noalign{\smallskip}
\hline
\multicolumn{3}{c}{} \\
$P_{orb}$ = 1.596 ~${\pm}$~ 0.003 days	& &	$e = 0.18 ~\pm~ 0.08$				\\
$T_{0}$(HJD)= 2~452~097.03 ~${\pm}$~ 0.17 	& &	$i = 0.77\degr ~\pm~ 0.09\degr$		\\
$K_{2}$(abs) = 2.4 ~${\pm}$~ 0.2 km s$^{-1}$	& &	$K_{1}$(emiss) = 2.7 ~${\pm}$~ 0.2 km s$^{-1}$		\\
$\gamma$(HC) = +4.5 ~${\pm}$~ 0.2 km s$^{-1}$		& &	$\omega$ = + 283$\degr$ ~${\pm}$~ 28$\degr$	\\
$d = 1.3 \pm 0.2$ kpc 		& &	$E(B-V)$ = 0.155~${\pm}$~0.007			\\
$q = M_{2}/M_{1} = 1.13~\pm~0.11$		& &							\\
\multicolumn{3}{c}{} \\
\hline
\noalign{\smallskip}
\multicolumn{3}{c}{Parameters of the primary star} \\
\noalign{\smallskip}
\hline
\multicolumn{3}{c}{} \\
$P_{1}$ = 0.205 days			& &	$M_{1}$ = 4.2  ~${\pm}$~ 0.7$M_{\sun}$	\\
$K$(0.205 d) = 2.3 ~${\pm}$~ 0.3 km s$^{-1}$	& &	$M_{B} = -0.6$  \\
$T_{1}$(HJD) = 2~452~299.62 ~${\pm}$~ 0.01		& &	$a_{1}$ = 6.3 ~${\pm}$~ 0.6 $R_{\sun}$         	\\
$P_{2}$ = 0.64 days			& &	                  	\\
$K$(0.64 d) = 1.1 ~${\pm}$~ 0.4 km s$^{-1}$	& &	    \\
$T_{2}$(HJD) = 2~452~097.40 ~${\pm}$~ 0.06		& &	 \\
\multicolumn{3}{c}{} \\
\hline
\noalign{\smallskip}
\multicolumn{3}{c}{Parameters of the secondary star} \\
\noalign{\smallskip}
\hline
\multicolumn{3}{c}{} \\
$M_{2}$ = 4.8~${\pm}$~ 0.5 $M_{\sun}$		& &	$a_{2}$ = 5.5 ~${\pm}$~ 0.5 $R_{\sun}$	\\
$R_{2}$ = 3.4 $R_{\sun}$			& &	log g = 4.05                       \\
$T_\mathrm{eff2}$ = 13~500 K			& &	Spectral type = B7$\pm$1~V          \\
$R_{r2}$(min) = 3.7~${\pm}$~ 0.8 $R_{\sun}$	& &	$R_{r2}$(max) = 5.4 ~${\pm}~ 1.1~ R_{\sun}$\\
$M_{B} = -0.7$		&	&	\\
\multicolumn{3}{c}{} \\
\hline
\end{tabular}
\end{flushleft}
\smallskip\noindent
\end{table*}

\subsection{The distance to the system}

       Heap \& Aller (\cite{heap}), as well as WS83 determined the distance to the system
based on the secondary star spectroscopic parallax. The first authors found a distance
of 1.26 kpc. They also determined that the neighboring star \object{HD~44498} has a reddening
of $E(B-V) = 0.14$. WS83 determined a distance of 1.208 kpc. The color excess
measured by them is $E(B-V) = 0.15$ and it is quite reliable, given that it is based on the
interstellar extinction band at 2200 {\AA}.

      In order to obtain an independent estimate of the interstellar extinction, we
measured the equivalent width of the diffuse interstellar band (DIB) at 5780 {\AA} and found W$_\lambda$(5780 {\AA})$ = 100 \pm 6$ m{\AA}.
Using the relation of the DIB equivalent width versus $E(B-V)$ color excess (Somerville \cite{some})

\eq
E(B-V) = [W_\lambda(5780 {\AA}) + 46]/940 
\eeq

\noindent      where W$_\lambda$ is given in m{\AA}, we obtain, $E(B-V) = 0.155 \pm 0.007$. The estimated
error does not take into account the uncertainty of Somerville's calibration. The two
methods of estimating the color excess agree quite well. One should notice, however,
that the DIB at 5780 {\AA} may be contaminated by the \ion{N}{iv} 5776.3 {\AA} emission line.

      We measured the equivalent width of \ion{Mg}{ii} 4481 {\AA} from the FEROS spectra of two comparison stars and
obtained for \object{HD~103401} (B5 V), W$_\lambda$(4481 {\AA}) = 0.263 $\pm$ 0.005 {\AA} and for  \object{HD~102465} (B9 V),
W$_\lambda$(4481~{\AA})~=~0.377 $\pm$ 0.004 {\AA}.  Interpolating the spectral type one would expect, for
the B7$\pm$1~V secondary component of \object{HD~45166},
W$_\lambda$(4481 {\AA}) = 0.320 $\pm$ 0.005 {\AA}. We found 0.177 $\pm$ 0.005 {\AA}, instead, due to the
dilution from the companion's continuum. This means that the secondary star
contributes 55\% $\pm$ 2\% in the $B$ filter, and the distance to \object{HD~45166} is, therefore, d = 1.3 $\pm$ 0.2 kpc.

	The radial velocity of the interstellar \ion{Na}{i} lines determined by us (see Table~\ref{tababs})
is 22.8 km s$^{-1}$ in the heliocentric system. After correcting this value to the Local Standard
of Rest (LSR) we obtain a radial velocity of V$_{LSR} = +7.1$ km s$^{-1}$. From the galactic
rotation curve we obtain, for the coordinates of \object{HD~45166} (l = 203.1$\degr$ / b = -2.3$\degr$), the expected radial
velocity of 10.8 (d/kpc) km s$^{-1}$. 
The radial velocity of the average stellar spectrum, corrected for the gravitational
redshift, is V$_{LSR} = -9.0$ km s$^{-1}$. We conclude, therefore, that the star has an anomalous
velocity of approximately $-23.1$ km s$^{-1}$ with respect to LSR.

\subsection{DACs and the photospheric connection}

	The coincidence between the orbital period ($P=1.596 \pm 0.003$ days) and the recurrence time of $1.60 \pm 0.15$ days (WHSH89) for the
Discrete Absorption Components (DACs) is striking.
 Why should exist an
association between the recurrent appearance of slowly accelerating clouds and the orbital period? Somehow the clouds responsible for the periodic
absorption seem to be ejected at every orbit. What could be the periodic event,
responsible for this ejection? At a first glance one could imagine that the non-zero
eccentricity may play a key role in this game. At every orbit there is a periastron event in
which the secondary fills its Roche lobe and is capable of transferring matter to the
primary. This periodic accretion may be physically related to the DACs but how, exactly, is not
clear at this point. We will revisit this issue in Paper II.

\section{Conclusions} \label{secc}

	In what follows we present the main conclusions of this paper:

\begin{enumerate}

\item The optical spectrum of \object{HD~45166} presents a great wealth of information both in
the emission as well as in the absorption spectrum. The emission spectrum has
lines of H, \ion{He}{i/ii}, \ion{C}{iii/iv}, \ion{N}{iii/iv/v}, \ion{O}{ii/iii/iv}, \ion{Si}{iii/iv}. The spectrum in absorption has lines of H, \ion{He}{i}, \ion{C}{ii}, \ion{N}{i}, \ion{O}{i},
\ion{Si}{ii}, \ion{Mg}{ii} and \ion{Fe}{ii}. Weak lines of \ion{C}{i}, \ion{Mg}{i}, \ion{Si}{i}, \ion{P}{ii}, \ion{S}{ii}, \ion{Cr}{ii} and \ion{Fe}{i}  are also
seen.

\item We classified the spectral type of the secondary star as B7~V and, therefore,
it should have a mass of M$_2$ = 4.8 M$_{\sun}$ and a radius of R$_2$ = 3.4 R$_{\sun}$.

\item The emission lines have great diversity of widths and profiles. The most intense
lines have FWHM of 370 km s$^{-1}$ and the weakest lines of 70 km s$^{-1}$. Lines of H and
He have Voigt/Gauss profiles and are systematically broader than the lines of
CNO, that have Lorentz profiles.

\item \object{HD~45166} is shown to be a double spectroscopic binary with orbital period of 1.596
$\pm$ 0.003 days and eccentricity of e = 0.18 $\pm$ 0.08.

\item Standard techniques for period search applied to the emission lines show two
additional periods, of 5 and 15 hours. 

\item The amplitude of the radial velocities of the orbital period are K$_1$ = 2.7 $\pm$ 0.2 km s$^{-1}$
and K$_2$ = 2.4 $\pm$ 0.2 km s$^{-1}$. We derived M$_1$ = 4.2 $\pm$ 0.7 M$_{\sun}$ and i = 0.77$\degr$ $\pm$
0.09$\degr$.

\item  The secondary star's radius may be about the size of its Roche lobe at
periastron. Therefore we can consider that, at periastron, it fills or it is very close to
fill its Roche lobe.

\item  The expected gravitational redshift is 0.9 km s$^{-1}$ for the secondary star and 2.0 km s$^{-1}$
for the primary. These values are larger than the precision of our measurements
and they need to be taken into account in the analysis of the data.

\item  We estimated a color excess of $E(B-V)=0.155~\pm~0.007$ and the distance to \object{HD~45166} as 1.3 $\pm$ 0.2 kpc.

\item  We suggest that the discrete absorption components (DACs) observed in the
ultraviolet with a periodicity similar to the orbital period may be induced by the
periastron events.

\end{enumerate}

\begin{acknowledgements}
We would like to thank Dr. Albert Bruch for his careful reading of the manuscript and the anonymous referee
for the constructive comments. 
We are specially grateful to Dr. Luiz Paulo Vaz for kindly fitting the
measurements of the radial velocity with his program of orbit adjustment with non-zero
eccentricity.
\end{acknowledgements}


\end{document}